\documentclass[]{emulateapj}
\shorttitle{Detailed Analysis of Balmer Lines in AGN}
\shortauthors{La Mura et al.}

\begin{document}

\title{Detailed Analysis of Balmer Lines in a SDSS Sample of 90 Broad Line AGN}
\author{G. La Mura\altaffilmark{1}, L. \v C. Popovi\'c\altaffilmark{2,3},
S. Ciroi\altaffilmark{1}, P. Rafanelli\altaffilmark{1}, D. Ili\'c\altaffilmark{4}}

\altaffiltext{1}
{
Department of Astronomy, University of Padova, Vicolo
dell'Osservatorio, I-35122 Padova, Italy;
giovanni.lamura@unipd.it, stefano.ciroi@unipd.it,
piero.rafanelli@unipd.it
}
\altaffiltext{2}
{
Astronomical Observatory, Volgina 7, 11160 Belgrade
74, Serbia; lpopovic@aob.bg.ac.yu
}
\altaffiltext{3}
{
Isaac Newton Institute of Chile, Yugoslavia Branch
}
\altaffiltext{4}
{Department of Astronomy, Faculty of Mathematics, University of Belgrade, 
Studentski trg 16, 11000 Belgrade, Serbia; dilic@matf.bg.ac.yu}

\begin{abstract}
In order to contribute to the general effort aiming at the improvement of our knowledge about
the physical conditions within the Broad Line Region (BLR) of Active Galactic Nuclei (AGN),
here we present the results achieved by our analysis of the spectral properties of a sample
of 90 broad line emitting sources, collected at the Sloan Digital Sky Survey (SDSS) database.
By focusing our attention mainly onto the Balmer series of hydrogen emission lines, which
is the dominant feature in the optical wavelength range of many BLR spectra, we extracted
several flux and profile measurements, which we related to other source properties, such
as optical continuum luminosities, inferred black hole masses, and accretion rates. Using
the {\it Boltzmann Plot} method to investigate the Balmer line flux ratios as a function
of the line profiles, we found that broader line emitting AGN typically have larger
$\rm H\alpha / H\beta$ and smaller $\rm H\gamma / H\beta$ and $\rm H\delta / H\beta$ line
ratios. With the help of some recent investigations, we model the structure of the BLR and
we study the influence of the accretion process on the properties of the BLR plasma.
\end{abstract}


\section{Introduction}
The dominant features in many Active Galactic Nuclei (AGN) spectra are broad emission lines which
originate in the Broad Line Region (BLR) \citep{osterbrock89, krolik99, peterson03}. The
BLR can potentially provide a useful probe of the central source, thus understanding its physics
and kinematics is a crucial step in the investigation of AGN. There are three reasons: (i) the
kinematics of the BLR is probably controlled by the central source, with the competing effects of
gravity and radiation pressure; (ii) the BLR reprocesses the X-ray/UV energy emitted by the
continuum source, consequently the broad emission lines can provide indirect information about
this part of the spectrum; (iii) there is indication that the parameters of the broad lines
(coming from the BLR) may be related to other fundamental properties of the source. Most of the
recent BLR studies have been focused on geometries, sizes, and correlations between the BLR
kinematical properties and the general characteristics of AGN \citep[see e. g.][etc.]{sulentic00,
popea04, kaspi05}, while other works have been devoted to reconstructing the physical conditions in
the BLR emitting gas \citep[etc.]{kaspi99, popovic03, popovic06, korista04, ver06}.

The broad line strength, width, and shape are powerful tools for gas diagnostics in the different
parts of the emitting region of AGN \citep[e. g.][]{osterbrock89}. However, there is the problem that
the broad emission lines are complex and that they are probably coming from at least two regions
with different kinematical and physical conditions. Furthermore, the broad emission line profiles
of some AGN may be explained with a two-component model \citep[see e. g.][]{popea02, popea03,
popea04, ilic06, bon06}.

Here we describe an extensive investigation of the spectral properties of the broad components
in the Balmer emission lines, combined with the study of those other source physical parameters
that we were able to infer from our sample of spectra. Exploiting some recent results that have
been achieved by means of the {\it Reverberation Mapping} (RM) technique \citep[cfr.][]{kaspi05,
bentz06}, we try to determine the central source masses and luminosities throughout the sample
and to costrain their role in controlling the BLR structure and dynamics. 

The paper is organized according to the following plan: in \S~2 we describe the selection of our
sample and we summarize the required reduction process before performing our measurements;
in \S~3 we report the extraction of our measurments and their related uncertainties; \S~4
will list our results, with a discussion of our findings and their limits, while in \S~5 our
conclusions are given.

\section{Sample selection and processing}
The set of spectra for our data sample (see an example in Fig.~\ref{f01}), has been collected at the
spectral database of the $3^{rd}$ data release from the SDSS. Observations are performed as survey
campaigns at the 2.5m $f / 5$ modified Ritchey-Chretien altitude-azimuth telescope, located
at the Apache Point Observatory, New Mexico (USA)\footnote{http://www.sdss.org/dr3}. Data
are obtained with a spectrograph, whose sensor uses a mosaic made up by four SITe/Tektronix 2048
x 2048 CCDs, which covers a wavelength range running from 3800\AA\ to 9200\AA. A
system of 640 optical fibers, each having an aperture of 3$''$, subdivides the telescope's field
of view, so that each exposure yields 640 spectra corresponding to as many areas in the sky. The
spectral resolution {\it R} of the observations ranges from 1850 to 2200. According to the
purposes of our work, we searched the SDSS database for sources corresponding to the following
requirements:
\begin{figure}[t]
\includegraphics[width = 8.2cm]{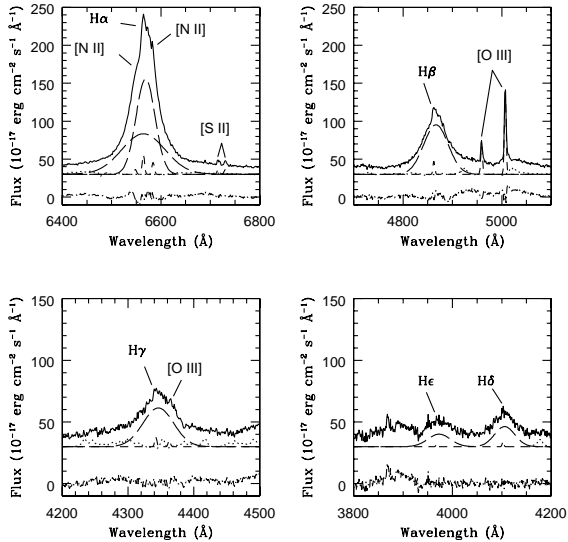}
\caption{Multiple Gaussian decomposition of the Balmer line profiles in the spectrum of
SDSS~J1203+0229. Here we denote with the appropriate labels all the emission lines that we
took into account. The observed spectrum (thick continuous line) has been artificially
shifted upwards for a better comparison with the narrow (short dashed), broad (long dashed),
and \ion{Fe}{2} (dotted) contributions. The dot-dashed line in the bottom part of each panel
gives the fit residuals.\label{f01}}
\end{figure}
\begin{enumerate}
\item objects had to be located at redshift $z < 0.4$, in particular it was required that the entire
profiles of lines belonging to the Balmer series were covered by the available spectral range;
\item only spectra where the Balmer series was clearly recognized, at least up to the $\rm H\delta$,
have been considered;
\item the profile of a broad component had to be detectable for each Balmer line;
\item such profiles had not to be affected by distortions, due, for example, to bad pixels on
the sensors, as well as by the presence of strong foreground or background sources.
\end{enumerate}

The preview spectra provided by the database retrieval software were manually inspected,
looking for the objects in better agreement with our requirements, until a number of 115 sources
were chosen from approximately 600 candidates examined in various survey areas. Subsequent
inspection of the spectra collected within the database led to the rejection of 25 objects, which
were affected by problems that could not be detected in the preview analysis. Therefore our
resulting sample includes the spectra of 90 variously broad line emitting AGN, corresponding to
$\sim 15\%$ of the candidates that we examined and located in the range $0.024 \leq z \leq 0.368$,
with an average redshift of 0.119. A significant fraction of our sample (45 spectra) has been
selected from the collection of sources studied by \citet{boroson03}. 

The SDSS database provides users with pre-processed material, therefore spectra retrieved from
the survey are already corrected for instrumental and environmental effects, including the
sky emission subtraction and the correction for telluric absorptions. Calibration of data
in physical units of flux and wavelength is also performed. Consequently, our preliminary
reduction simply needs to take into account a correction for Galactic Extinction, which we
estimated using an empirical selective extinction function \citep[see][]{cardelli89} that
was computed for each spectrum on the basis of the Galactic Extinction coefficients given
by \citet{schlegel98} and available at the {\it Nasa Extragalactic Database} ({\it
NED})\footnote{http://nedwww.ipac.caltech.edu/}, and the removal of cosmological redshift.
Since our interest lies on the investigation of the BLR properties, we had to identify the
broad line components in spectra. In principle, the BLR signature can be isolated if we know
which contributions are introduced by the underlying continuum of both the AGN and its host
and by the Narrow Line Region (NLR). In many cases large samples are dealt with by means of automatic
data processing techniques, which can be particularly useful in a statistical sense, but they
neglect most of the peculiar properties of the single sources. In our case, intrinsic differences
among the various objects certainly did not simplify the task. Broad and narrow lines,
indeed, usually have blended profiles, whose final shape may critically depend on many
circumstances, such as the orientation of our line of sight onto the source, the presence of
absorbing material, or the amount of signal originating outside the source itself. For this reason
we undertook the task of manually identifying the BLR contributions in spectra, rather than relying
on automatically collected determinations. We used the IRAF software for reduction and analysis
of our data. At first, we performed a continuum normalization, fitting the underlying continuum
shape of each spectrum in the rest frame wavelength ranges typically running between 3750 - 3850\AA,
4200 - 4250\AA, 4700 - 4800\AA, 5075 - 5125\AA, 5600 - 5700\AA, 6150 - 6250\AA, and 6800 - 7000\AA,
which were not affected by significant line contamination. The shape of the spectral continuum was
reproduced by means of {\tt spline} functions with order ranging from 3 to 7 in the {\tt splot} task
of IRAF. As shown in Fig.~\ref{f01}, subtraction of the narrow line components has been achieved by
means of multiple Gaussian fits to their profiles. Using the narrow [\ion{O}{3}] feature at 5007\AA,
we extracted a template profile for forbidden lines in each spectrum and the other narrow lines were
required to have compatible widths. The line intensity ratio for the [\ion{N}{2}] doublet at $\lambda
\lambda$ 6548, 6584 has been fixed to $1 : 3$. In a certain number of cases, before proceeding with
our measurements of fluxes and profiles, we had to remove the contribution due to the \ion{Fe}{2}
multiplets. According to the suggestions given by \citet{ver04}, the removal of spectral
contributions from \ion{Fe}{2} can be performed by scaling and smoothing a template \ion{Fe}{2}
spectrum,  which was previously extracted from the spectrum of {\it I Zwicky 1} \citep{botte04}.
We scaled the template according to the \ion{Fe}{2} features seen in spectra and we measured the
resulting fluxes in order to estimate the \ion{Fe}{2} contribution. However this step was
unnecessary for many objects in the sample, since the \ion{Fe}{2} emission lines gave negligible
contributions to the fluxes or they were undiscernible above the average noise intensity fluctuations
in these sources.

\subsection{Host galaxy correction}
\begin{figure}[t]
\includegraphics[width = 8.2cm]{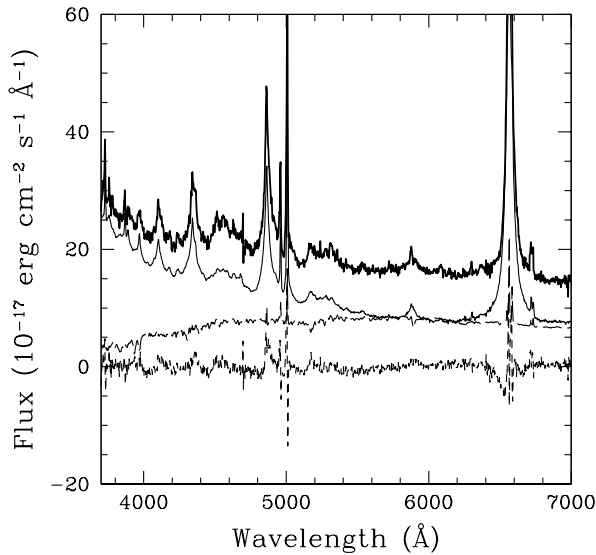}
\caption{Spectral decomposition of SDSS J1307--0036. In this plot we show the spectrum of the
object in the rest frame wavelength range used to perform the fit (thick continuous line),
the AGN component (thin continuous line), the host galaxy contribution (long dashed line),
and the fit residuals (short dashed line).\label{f02}}
\end{figure}
\begin{figure}[t]
\includegraphics[width = 8.2cm]{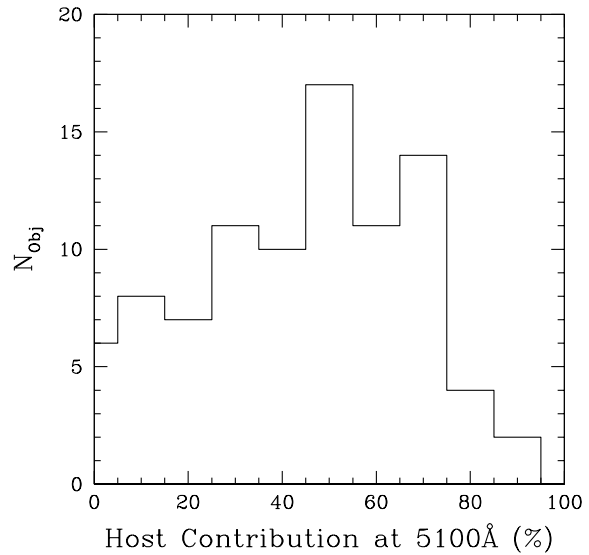}
\caption{Relative importance of host galaxy contamination to the optical continuum in the spectra
of our sample. The distribution does not show any particular dependence on the object redshift or
its luminosity.\label{f03}}
\end{figure}

Light from targets observed by the SDSS spectrographs is collected within sky areas of fixed
aperture, which, in the case of AGN, will bring to various contributions from the host galaxy to
the total flux recorded by instruments, essentially depending on the object redshift and on the
relative importance of the host with respect to its AGN. A possible way to account for the
influence of host galaxies on the formation of the resulting spectra is the one proposed by
\citet{vdb06} on the basis of the {\it Karhunen-Lo\`eve Transforms} described in \citet{conn95}.
This technique assumes that the total spectrum of an AGN and its host galaxy may be the result of
a sum of orthonormal components, which make up a set of eigenspectra, arranged in a linear
combination such as:
$$S(\lambda) = \sum_{i = 1}^{n}[a_i\cdot A_i(\lambda)] + \sum_{j = 1}^{m}[h_j\cdot H_j(\lambda)],
\eqno(1)$$
where we call $S(\lambda)$, $A_i(\lambda)$, and $H_j(\lambda)$ the total spectrum, the $i^{th}$
AGN eigenspectrum, and the $j^{th}$ host galaxy eigenspectrum, while $a_i$ and $h_j$ are
respectively the AGN and host galaxy eigencoefficients for the corresponding components, which
do not depend on wavelength. The possibility to estimate the host galaxy contribution in the
resulting spectrum stems from the use of appropriate sets of eigenspectra, like those computed
by \citet{yipa04, yipb04}, both for galaxies and AGN, from large samples of galaxy and AGN SDSS
spectra. We used an iterative $\chi^2$ minimization process, which was run over the rest frame
wavelengths covering the range from 3700\AA\ to 7000\AA, to assign the appropriate
eigencoefficient values for the decomposition of our spectra. In a similar analysis to the
one performed by \citet{vdb06}, we found that most of the variations within our sample could
be accounted for by introducing five galaxy and six AGN components, because a smaller number
of eigenspectra left very large residuals, whilst a larger number took to a dangerous
overfitting of noise fluctuations. The decomposition of our spectra usually yielded a
reduced residual of $0.05 < \chi^2 < 2$ in the continuum. In Fig.~\ref{f02} we give an example
of a spectral decomposition obtained by combining separately all the AGN and host components.
The figure illustrates that this method is actually able to fit the spectral continuum in good
detail, although significant residuals are left in the regions corresponding to the line cores,
where the highest order variations of our sample are carried out. The spectral contributions
identified as being due to the host were therefore removed from our subsequent measurments of
the AGN properties and they have been used to estimate the relative importance of the AGN and
its host in the emission of the observed luminosities. Fig.~\ref{f03} illustrates the amount
of host galaxy contaminations detected in our sample in the form of a histogram. When we
studied the distribution of these non-AGN components, we found no clear evidence for a redshift
or luminosity dependence. Indeed, taking into account objects at larger distances, there is a
higher probability to include more powerful AGN and this may balance the effect of light
collected from a wider fraction of the host.

\section{Spectral analysis}
Once the broad line components had been fairly isolated in the spectra, we performed a
number of measurements in order to estimate fluxes, FWHM, and FWZI. The fluxes of the
Balmer lines were measured several times and the uncertainties to be associated to
the line flux determinations were computed as:
$$\sigma_F = \sqrt{\left(\frac{\sigma_{\rm cont}}{I_{50}}\cdot F\right)^2
+ {\bar{\sigma}_F}^2}, \eqno(2)$$
where $\sigma_{\rm cont}$ represents the rms due to noise fluctuations estimated in the continuum
close to the emission lines, $F$ and $I_{50}$ are, respectively, the total fluxes of the lines and
their half maximum intensities, while $\bar{\sigma}_F$ is the standard deviation of our multiple
flux determinations with different choices of the continuum intensity level. A list of flux ratios
for various Balmer lines, with respect to $\rm H\beta$ from different sources, is given in
Table~\ref{t01}. Their error bars are obtained as:
$$\sigma_{Rj} = R_j \cdot \sqrt{\left(\frac{\sigma_{Fj}}{F_j}\right)^2
+ \left(\frac{\sigma_{H\beta}}{F_{\rm H\beta}}\right)^2}, \eqno(3)$$
where $R_j$ is the ratio of $j^{th}$ Balmer line with respect to $\rm H\beta$. Similar flux
measurments have been performed on the [\ion{O}{3}] emission line at 5007\AA\ and they are
also reported in Table~\ref{t01}.

In order to estimate the FWHM and FWZI, we restricted our attention only on the strongest
spectral features of the BLR, namely $\rm H\alpha$ and $\rm H\beta$, and the [\ion{O}{3}] 
5007 line of the NLR. After choosing a zero emission intensity level, we computed the line peak
intensities and we used the inferred values to define a half maximum intensity. Half widths
at half the maximum and at zero intensity were taken both on the red and the blue line wings,
respectively where the line profiles crossed the half and the zero intensity levels. Such
determinations were repeated several times for each line, taking into account different guesses
to the zero intensity levels, and they have been eventually averaged together, thus providing
the mean values with their standard deviations. We corrected the emission line profiles for
instrumental broadening, assuming that the profiles were affected according to:
$$W_{\rm obs} = \sqrt{W_{\rm int}^2 + W_{\rm ins}^2}, \eqno(4)$$
where $W_{\rm obs}$ and $W_{\rm int}$ are the observed and intrinsic line widths, while
$W_{\rm ins}$ characterizes the instrumental broadening, which, in terms of velocity units,
may be expressed as $W_{\rm ins} = c / R \approx 150 {\rm km\ s^{-1}}$. 

To estimate the continuum luminosities, we performed averaged flux measurements over the
rest frame wavelength range running from 5075\AA\ to 5125\AA. After the correction
for Galactic Extinction had been taken into account, the fluxes, that we derived both in the
continuum and in the selected emission lines, were converted into specific luminosities, making
the assumption of isotropic emission of radiation from the sources. Here we used the cosmological
redshift as a distance estimator, in the framework of a model characterized by $H_0 = 75 {\rm
km\ s^{-1}\ Mpc^{-1}}$, $\Omega_m = 0.3$, and $\Omega_\Lambda = 0.7$. We assumed the sources
included in our sample to have optical luminosities given by $L_{5100} = 5100\cdot L_\lambda$,
where $L_\lambda$ is the specific continuum luminosity measured at 5100\AA, and their bolometric
luminosities, which we needed in order to guess the accretion rate onto the central source, to
be roughly ten times as much.

\section{Results}
Since we are going to consider the influence of the central source of an AGN onto the BLR, we
must necessarily confront the problems due to its currently unresolvable structure. The
conversion of observational data into physical parameters requires some assumptions to be made
about the BLR structure and dynamics. At present, the most trusted interpretation of
the broad emission line profiles, which we can often detect in AGN spectra, invokes fast
orbital motions of a photoionized plasma in the gravitational field of a Super Massive Black
Hole (SBH). Matter accretion onto the black hole provides an energetic continuum of radiation
that interacts with the gas distribution surrounding the central source and it eventually results
in the observed spectra. Concerning the real behaviour of such gas distribution, there are many
aspects which still have to be unambiguously clarified. Recent investigations \citep[e. g.][]
{vestergaard00, nikolajuk05} suggest that the BLR geometry should be considerably flattened,
rather than spherical, and its motions should occur mainly in an orbital configuration, as
opposed to radial infalls or outflows.

Given that the BLR is dynamically affected by the gravitational field of the central source
and by its radiation pressure, the broad line emitting medium must be controlled by the balance
of these forces. As a consequence, the mass of the central engine and its energetic output should
be tightly related to the size and velocity field measured in the BLR \citep{wandel99, wu04,
peterson04, kaspi05}. In order to estimate the size of the BLR, a number of AGN have been studied
with the {\it Reverberation Mapping} (RM) technique, which exploits the time lags elapsed since
the ionizing continuum variations and the corresponding line responses to map the radial
distribution of the line emitting gas. Although this method is currently one of the most powerful
available probes of the structure of the BLR, it is very expensive, because it requires large
monitoring campaigns for each object under study. Using the available data it is
nonetheless possible to investigate how the BLR radii change as a function of the AGN
luminosities in various spectral ranges. In the optical domain \citet{kaspi05} argued that the
size of the BLR scales with the source luminosity according to:
$$\frac{R_{\rm BLR}}{10 \mbox{lt-days}} = (2.23 \pm 0.21) \left[\frac{\lambda L_\lambda(5100
\mbox{\AA})} {10^{44} {\rm erg\cdot s^{-1}}}\right]^{0.69 \pm 0.05}. \eqno(5)$$
More recently \citep[see e. g.][]{bentz06}, this kind of relationship has been reviewed for
different RM sources and it has been suggested that the power law index of Eq.~(5) may be
consistent with the value of $\alpha = 0.5$, predicted on the basis of some simple
photoionization calculations. In the following we adopt the latter interpretation to
infer the source parameters from our measurments.

Assuming the BLR to be in virial equilibrium \citep[see e. g.][and references therein, for a
discussion of implications]{woo02, sulentic06} and to have an isotropic motion pattern, the
latter approximation being forced by the lack of multi-wavelength information about the real
flattening of the BLR and its inclination with respect to our line of sight \citep[see for
example][for a discussion of the BLR inclination]{bian05}, we estimate the black hole masses
by means of:
$$M_{\rm BH} = \frac{R_{\rm BLR}}{G}\left(\frac{\sqrt{3}}{2} \mbox{FWHM}_{\rm H\beta}
\right)^2. \eqno(6)$$
Given the optical luminosities and the inferred black hole masses we estimate the accretion
rates onto the central mass with:
$$\dot{M} = \frac{L_{\rm Bol}}{\eta c^2} = \dot{M}_{\rm Edd} \frac{L_{\rm Bol}}{L_{\rm Edd}},
\eqno(7)$$
assuming that the source bolometric luminosities can be roughly expressed as $L_{\rm Bol}
\approx 10 L_{5100}$ and that the efficiency of the accretion process is $\eta = 0.1$.
\begin{figure}
\includegraphics[width = 8.2cm]{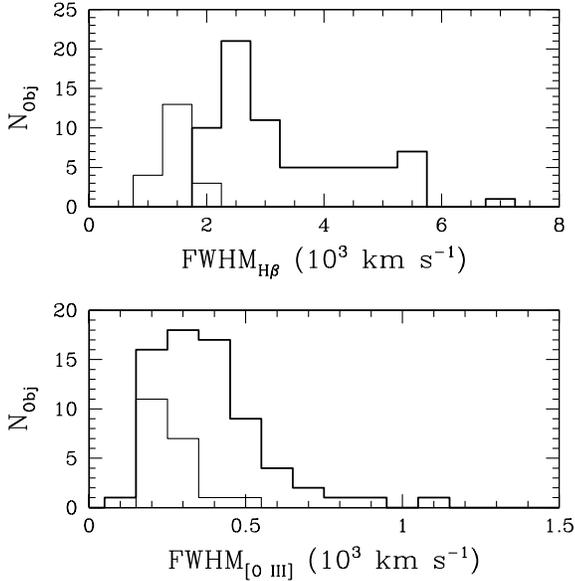}
\caption{Line width distribution of the sample, plotted as a comparison of the FWHM in $\rm
H\beta$ (top panel) and the corresponding value in [\ion{O}{3}] $\lambda 5007$ (bottom panel). The
thin continuous histograms show the distribution of NLS1s with $\mbox{FWHM}_{\rm H\beta} < 2000
{\rm km\ s^{-1}}$, while the thick continuous histograms represent broad line emitting AGN.
\label{f04}}
\end{figure}

In the framework of these assumptions we find that our sample covers a black hole mass range
between $5.59\cdot 10^5 {\rm M_\odot}$ and $2.96\cdot 10^8 {\rm M_\odot}$, with the inferred
accretion rates running from 0.024 to 0.835. We give a summary of the results achieved by our
calculations in Table~2. Though most of our estimates to the physical and structural properties
of the BLR in these objects have been inferred by means of simplified assumptions and empirical
relations, so that they might in principle be prone to large uncertainties, the sample, as a whole,
provides a useful reference frame, where the actual properties of realistic sources can be expected
to span. Since the selection of our sample is limited in redshift by the requirement of the Balmer
series falling into the available spectral range, it lacks the most powerful broad line AGN.
However, as we show in Fig.~\ref{f04}, where we compare the FWHM of $\rm H\beta$ and the [\ion{O}{3}]
forbidden line, the sample appears to be fairly well distributed, although there are no objects
featuring very broad lines. Among the 90 spectra of the sample, 20 match the classical definition of
{\it Narrow Line Seyfert 1} (NLS1) galaxies, with $\mbox{FWHM}_{\rm H\beta} < 2000 {\rm km\ s^{-1}}$,
while the others are spread in the range of $2000{\rm km\ s^{-1}} < {\rm FWHM_{H\beta}} < 7000{\rm
km\ s^{-1}}$. NLS1s are typically different from other Seyfert galaxies under many points of
view. They are commonly observed to have super-solar metallicities in their nuclei and they often
show strong \ion{Fe}{2} multiplets, together with a soft slope and high variability in the X-rays
\citep[see e. g.][]{boller96}. Moreover they are usually observed to have relatively high 
Fe / [\ion{O}{3}] intensity ratios and strong soft X-ray excesses. Many explanations have been
proposed by various authors to give an interpretation of the properties found in NLS1s. Simply
assuming that the line broadening depends on the gravitational potential of the SBH and that the
BLR is not systematically larger in these sources, Boller et al. (1996) suggested that NLS1s are
powered by comparatively low mass black holes, with respect to normal Seyfert 1 galaxies. It has
also been proposed that they could be young AGN in a growing phase \citep{mathur01}, that the
narrow line profiles could be the effect of a partially obscured BLR \citep{smith02} or that they
could arise from an extremely flat geometry seen at low inclinations \citep{osterbrock85}, but the
polarization signatures which we should expect in the last two cases have not been reported so
far. Many of the characteristic properties of NLS1s could not be further investigated in our sample,
because of its limited spectral coverage and of the usually not ideal s/n ratios, which, in many
cases, did not allow us to actually measure the strength of \ion{Fe}{2} emission. In conclusion,
we found that NLS1s in our sample, located at an average redshift of $z = 0.064$, have on average
the $L_{\rm Fe II} / L_{5100} = 0.029 \pm 0.013$, $L_{\rm Fe II} / L_{\rm [O III]} = 4.30 \pm 3.08$,
and $L_{\rm Fe II} / L_{\rm H\beta} = 16.82 \pm 7.81$ luminosity ratios, while the other AGN have
$L_{\rm Fe II} / L_{5100} = 0.024 \pm 0.020$, $L_{\rm Fe II} / L_{\rm [O III]} = 4.36 \pm 5.31$, and
$L_{\rm Fe II} / L_{\rm H\beta} = 13.09 \pm 11.66$. In the following we kept the distinction of
NLS1s from the other AGN.

\subsection{The continuum source and Balmer Line properties}
\begin{figure}[t]
\includegraphics[width = 8.2cm]{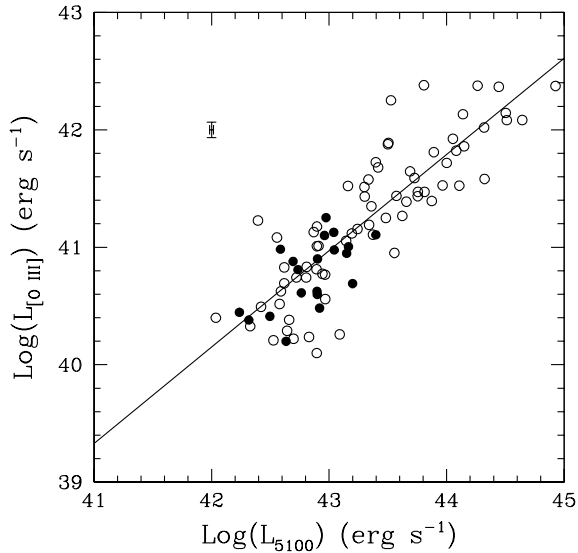}
\caption{Correlation between the [\ion{O}{3}] $\lambda 5007$ line luminosity and the optical
continuum luminosity measured in the spectra. NLS1s have been identified with filled
circles, while the other broad line emitting sources are represented by open circles. The
cross located in the upper left corner of the diagram shows the median uncertainty of our
measurements. The straight line fit is achieved by means of Eq.~(8). Intrinsic absorption
and starlight contributions might be responsible for the scatter.\label{f05}}
\end{figure}

At present it is strongly believed that the main interactions between the central power source
of an AGN and the surrounding emission line regions occurs via radiative processes. A direct
measurement of the AGN continuum luminosity on the spectra, however, is not straightforward
because of the several uncertainties about the actual AGN Spectral Energy Distributions (SED),
which can even be affected in various manners by stellar contributions from the AGN host
galaxy \citep[see][for instance]{bentz06}. On the other hand, some clues about the reliability
of the optical continuum intensity as an estimator of the ionizing radiation field strength
come from the observation of how the emission line intensities, which should be controlled by
the amount of ionizing radiation, correlate with the estimated continuum luminosities. In
Fig.~\ref{f05}, we show the distribution of our sample in the case of [\ion{O}{3}] $\lambda
5007$ as a function of the optical source luminosities. The behaviour that we find is
described by the relation:
$$\log(L_{\rm [O\ III]}) = (0.820 \pm 0.052)\cdot \log(L_{5100}) + (5.710 \pm 2.270), \eqno(8)$$
with a correlation coefficient $R = 0.712$ and a probability to occur by chance $P < 10^{-6}$.
Most of the scatter observed in this relation can be accounted for in terms of intrinsic
source absorptions or stellar light contributions which could have been not perfectly removed.

\begin{figure}[t]
\includegraphics[width = 8.2cm]{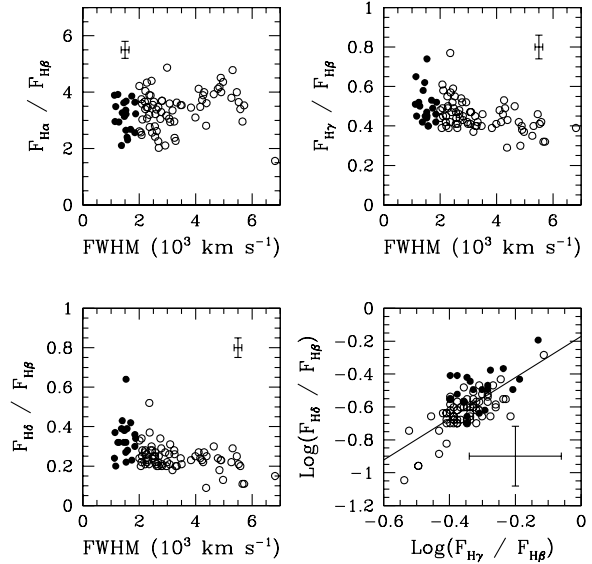}
\caption{Panels from the upper left to the lower left show the Balmer line flux ratios
with respect to $\rm H\beta$ as a function of the FWHM measured in spectra. We observe an
averagely larger Balmer decrement in objects with broader line profiles. The bottom right
panel illustrates the only correlation observed between $\rm H\delta / H\beta$ and $\rm
H\gamma / H\beta$. We plot with open circles the broad line emitting sources and with
filled ones NLS1 galaxies. The median uncertainty estimates are given by the crosses
shown in each plot.\label{f06}}
\end{figure}

Looking at the Balmer line intensity ratios over the sample, where we found the median
values of $\rm H\alpha / \rm H\beta = 3.45 \pm 0.65$, $\rm H\gamma / \rm H\beta = 0.45 \pm
0.08$, and $\rm H\delta / \rm H\beta = 0.25 \pm 0.08$, we compared the properties of the
sources with their associated line profiles. The plots shown in Fig.~\ref{f06} illustrate the
Balmer line flux ratios as a function of the FWHM measured in the spectra. The Balmer
line flux ratio measurements are prone to quite large uncertainties, presumably reflected by
the resulting scatter, but the plots suggest the existence of a weak relation between the Balmer
decrement and the width of line profiles. Although the correlations are not particularly
strong (having $R = 0.226$, $P = 3.34\cdot 10^{-2}$ in the case of $\rm H\alpha / \rm H\beta$,
while $R = -0.482$, $P = 1.71\cdot 10^{-6}$ for $\rm H\gamma / \rm H\beta$, and $R = -0.510$,
$P < 10^{-6}$ for $\rm H\delta / \rm H\beta$), they apparently strengthen with increasing line
widths, but this may be simply due to the small number of such objects. Almost no particular
correlations are, instead, observed among the Balmer line flux ratios themselves, with the
only exception of the ratios of $\rm H\delta / H\beta$ and $\rm H\gamma / H\beta$, whose
values are related by:
$$\log(F_{\rm H\delta} / F_{\rm H\beta}) = (1.253 \pm 0.133)\cdot \log(F_{\rm H\gamma} /
F_{\rm H\beta}) -$$
$$- (0.171 \pm 0.047), \eqno(9)$$
with a correlation strength of $R = 0.709$ and $P < 10^{-6}$. These findings agree very well
with earlier results obtained by \citet{raf85}, who, while investigating the relationships
among the Balmer line intensities and profiles in a sample of 12 Seyfert 1 galaxies,
recorded a correlation like the one given in Eq.~(9) (with a slope of 1.24) and the
absence of similar relations involving the other emission lines.

\subsection{Global Baldwin Effect}
\begin{figure}[t]
\includegraphics[width = 8.2cm]{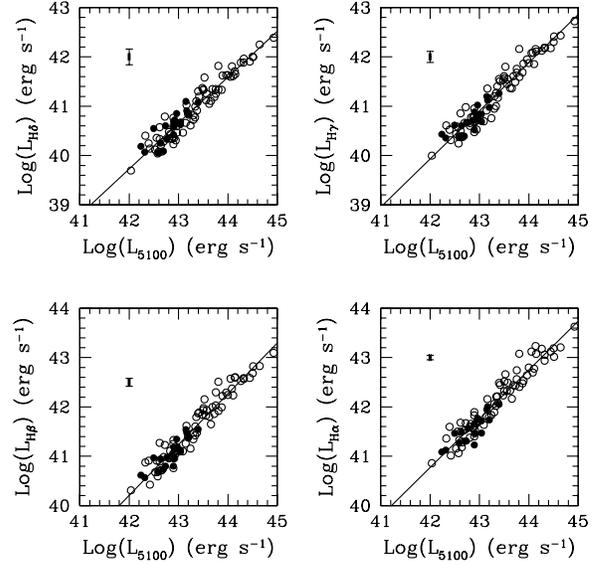}
\caption{Correlations between the optical continuum luminosity and the $\rm H\delta$
(upper left panel), $\rm H\gamma$ (upper right panel), $\rm H\beta$ (lower left panel),
and $\rm H\alpha$ emission line luminosities (lower right panel). Open circles represent
broad line emitting objects, while filled circles are NLS1s. The straight lines are the
best fits of Eq.~(10) and the error bars show the median uncertainties associated with
our measurements.\label{f07}}
\end{figure}

The occurrence of a power law slope like $L_{\rm [O\ III]} \propto {L_{5100}}^\alpha$
with $\alpha = 0.820 \pm 0.52$ in Eq.~(8) indicates the presence of a global Baldwin Effect
in this line \citep{bald77, kong06}. As we look for the situation of the broad Balmer line
components, however, we find little or no evidence for global Baldwin Effect, since the
observed emission line luminosities are related to the optical continuum through:
$$\log(L_{\rm H\alpha}) = (0.989 \pm 0.031)\cdot \log(L_{5100}) - (0.784 \pm 1.346) \eqno(10a)$$
$$\log(L_{\rm H\beta}) = (1.029 \pm 0.030)\cdot \log(L_{5100}) - (3.025 \pm 1.289) \eqno(10b)$$
$$\log(L_{\rm H\gamma}) = (0.976 \pm 0.029)\cdot \log(L_{5100}) - (1.063 \pm 1.226) \eqno(10c)$$
$$\log(L_{\rm H\delta}) = (0.931 \pm 0.030)\cdot \log(L_{5100}) + (0.631 \pm 1.299), \eqno(10d)$$
with correlation coefficients of $R = $ 0.937, 0.967, 0.974, and 0.972, spanning from
Eq.~(10a) to Eq.~(10d), and $P < 10^{-6}$ in all cases. The situation of the Balmer lines is
summarized in Fig.~\ref{f07}. Here we note that, although there are works which actually
detected a significant intrinsic Baldwin Effect, particularly in the case of $\rm H\beta$,
in long time covering observations of some well studied objects \citep{gilb03, goad04},
no strong indications have been found in literature concerning the existence of a
global Baldwin Effect in the broad Balmer lines \citep[see e. g.][]{yee80, bin93, osm99}.

\subsection{Balmer Decrements and Boltzmann Plots}
The broad line flux ratios estimated in our sample, which we presented in Fig.~\ref{f06},
are a potential probe to study how the physical properties of the BLR influence the Balmer
decrement. To investigate this effect we need to choose a specific parameter that
should be used as an estimator of the Balmer decrement and of its shape in each object. A
suitable possibility is to apply the {\it Boltzmann Plot} (BP) method \citep[see][for
description]{popovic03, popovic06} to the broad line components of the Balmer series.
Introducing a normalized line intensity with respect to the atomic constants involved in the
transition:
$$I_{\rm n} = \frac{F_{ul}\cdot \lambda_{ul}}{g_u\cdot A_{ul}}, \eqno(11)$$
where $F_{ul}$ is the measured flux, while $\lambda_{ul}$, $g_u$, and $A_{ul}$ are the line
wavelength, the upper level statistical weight, and the spontaneous radiative decay
coefficient respectively, it can be shown that the line intensities of a specific series of
transitions in an optically thin plasma depend on the excited level's energy according to:
$$I_{\rm n} \approx h c \cdot \ell \frac{N_0}{Z}\cdot \exp{\left(\frac{-E_u}{k_{\rm B}\cdot
T_{\rm e}}\right)}, \eqno(12)$$
in which we denoted with $N_0$ the number density of the radiating species, $Z$ the partition
function, $E_u$ the excited level energy, and $T_{\rm e}$ the excitation temperature, while
$\ell$ is the radial extension of the emitting region, and $h$, $k_{\rm B}$, and $c$ are the
fundamental constants of Planck, Boltzmann, and the speed of light. If the high excitation
stages (with $n > 2$) in hydrogen are well described by the Saha-Boltzmann distribution,
Eq.~(12) takes the form of:
$$\log{I_{\rm n}} = B - A\cdot E_u, \eqno(13)$$
where $A$ is called the {\it Temperature Parameter} because, in the situation depicted above,
it would be $A = \log{e} / (k_{\rm B} T_{\rm e}) \approx 5060 / T_{\rm e}\ [{\rm K^{-1}}]$.

\begin{figure}[t]
\includegraphics[width = 8.2cm]{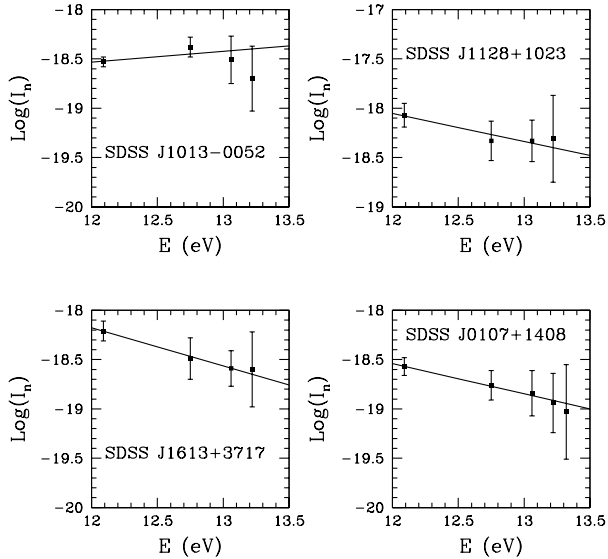}
\caption{Four examples of Boltzmann Plots applied to the spectra of objects belonging
to the sample: in the upper left panel no straight line fit is achieved (class iv);
in the upper right panel only a poor fit is performed (class iii); the bottom left
panel gives a good fit, but $\rm H\epsilon$ could not be detected in the spectrum (class
ii); finally the bottom right panel shows a straight line fit to the normalized
intensities of the Balmer series up to $\rm H\epsilon$ (class i).\label{f08}}
\end{figure}

\begin{figure}[t]
\includegraphics[width = 8.2cm]{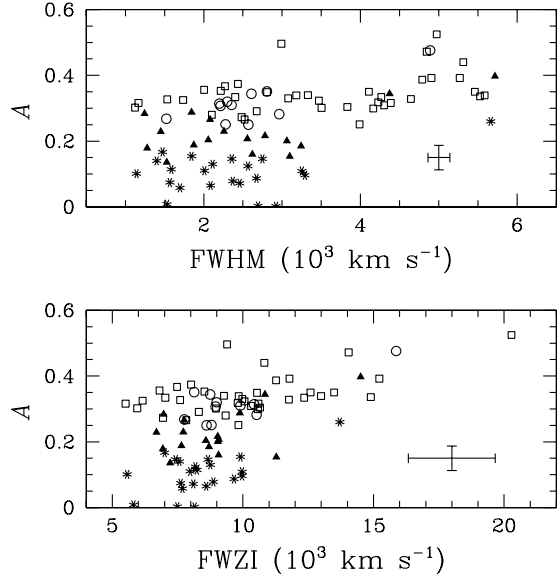}
\caption{BP slope $A$ as a function of the BLR velocity fields inferred by both the
FWHM and FWZI of the Balmer lines. Here we plot with open circles the sources where the
straight line fit covered the series up to $\rm H\epsilon$ (i) and with open squares those
objects having a fairly good fit up to $\rm H\delta$ (ii). Filled triangles are spectra with
a poor fit (iii) and finally asterisks are spectra where the Balmer line normalized
intensities could not be suitably described by a straight line (iv). The crosses in the
lower right corners of both panels are the median uncertainties of measurements.\label{f09}}
\end{figure}

When we apply the BP to the Balmer series of a BLR spectrum, the resulting slope $A$ clearly
depends on the Balmer decrement, though there is a substantial difference between these two
properties, since, for instance, we find objects with a rising BP slope ($A < 0$), although this
does not imply $\rm H\alpha$ to be really fainter than the other Balmer lines. The shape of the
Balmer decrement, too, influences the BP, which can be dramatically limited by the restrictions
involved in the assumptions about the physical conditions within the plasma. Intrinsic reddening
effects of the AGN environment, too, may influence the method, enhancing low order lines with
respect to the high order ones, thus taking to a steeper straight line fit. Therefore, in
presence of intrinsic reddening, the observed BP temperatures would be probably lower than the
actual values, although there are indications that the straight line fit to the normalized
intensities is not qualitatively affected \citep{popovic03}. According to whether the observed
spectra could be described by a relation like Eq.~(13), we studied the results of BP and we
found the occurence of four most common situations, corresponding to (i) a good fit of the
Balmer series up to the $\rm H\epsilon$ emission line, (ii) a good fit where, however, the $\rm
H\epsilon$ line could not be detected, (iii) a poor fit to the observed lines, and (iv) no
possible straight line fit of the line normalized intensities. Examples for each one of these
classes can be found in Fig.~\ref{f08}. We report the BP slopes and an indication of the proper
class in Table~1, while, in Fig.~\ref{f09}, we show the distribution of the inferred BP slopes
as a function of the BLR velocity fields. Here we see that the BP slope increases on average
with the broad line widths, as it was previously reported \citep{popovic03, popovic06}. The
total degree of correlation is quite weak ($R = 0.381$, $P = 2.27\cdot 10^{-4}$, in the case of
the relation between $A$ and FWHM, while it is $R = 0.470$, $P = 3.28\cdot 10^{-6}$ for that
between $A$ and FWZI), but it grows significantly if only the cases where a reasonable BP fit
is achieved are taken into account ($R = 0.453$, $P = 1.08\cdot 10^{-3}$ for $A$ and FWHM, $R =
0.579$, $P = 1.30\cdot 10^{-5}$ for $A$ and FWZI).

\subsection{Balmer decrements and Eddington ratios}
\begin{figure}[t]
\includegraphics[width = 8.2cm]{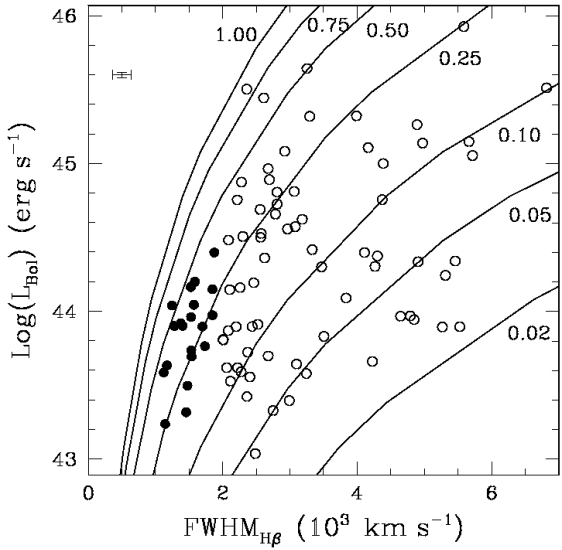}
\caption{Bolometric luminosities vs. FWHM$_{\rm H\beta}$ in the sample. Filled circles
represent NLS1s, while open circles are broad line AGN, and the cross in the upper left
corner gives the median uncertainty of the measurements. The curves track the expected
locations of black holes with masses between $10^4 M_\odot$ (bottom left side of the
diagram) and $10^9 M_\odot$ (upper right region) accreting at the labelled values of
$\dot{M} / \dot{M}_{\rm Edd}$, if their emission were isotropic, the accretion efficiencies
were $\eta = 0.1$, and the BLR scaled as described in \citet{bentz06}.
\label{f10}}
\end{figure}

The rate of the accretion process that produces the ionizing continuum of radiation in AGN is
a crucial parameter affecting the physical conditions of the emission line regions. Many
theoretical works suggest that the SED of an ionizing radiation field depends on the rate and
the efficiency of the accretion process \citep[e. g.][]{netzer92, kong06}, which, in turn,
affects the ionization status of the BLR and its stability \citep{nicastro00}. Unfortunately
the determinations of accretion rate, estimated from black hole mass and luminosity, may be
considerably affected by model dependent assumptions. In our case, the situation is shown in
Fig.~\ref{f10}, where we compare with our sample the expected distribution of AGN powered by
black holes with masses in the range $10^4 M_\odot < M_{\rm BH} < 10^9 M_\odot$ and accreting at
the labelled values of their Eddington ratios. We note that this model does not predict any
particular difference in the accretion rate of NLS1s and of the other sources. The accretion
rates that we infer by assuming a structural model computed according to \citet{bentz06} show
a rather small degree of correlation ($R = -0.314$, $P = 2.17\cdot 10^{-3}$) with the strength
of the Balmer decrement, as summarized in the BP slope $A$. As we illustrate in Fig.~\ref{f11},
this means that the high order lines appear to be stronger when the accretion rate is large,
with, perhaps, a flattening where the line ratios approach the expected atomic values. In
order to understand the relationship between the accretion rate and $A$, we should clarify the
role of this parameter. As mentioned above, in the appropriate circumstances the temperature
parameter $A$ is related to the thermodynamics of the emission line region. Although the
physical conditions of the BLR are such that a thermodynamic interpretation could not be
generally straightforward, it might be possible to explore some fraction of the BLR with this
technique. Especially in the objects where we found fairly good fits to the BP, the
dependence of $A$ on the line profiles and Eddington ratios could be a consequence of the
influence of ionizing radiation onto the emission line plasma. In this case the occurrence of
large temperature parameters in broad line emitting sources is an indication of low
ionization, maybe because of a shielding of the broad Balmer line emitting gas, while the
anti-correlation between the Balmer decrement and the Eddington ratio might be a consequence
of high dust column densities in our sightline towards the objects with low accretion rate.
It is possible that the temperature parameter $A$ tracks the influence of the ionizing
radiation field onto the broad line emitting plasma, therefore its relations with the line
profiles and Eddington ratios would suggest a stronger ionization in narrow line and in
high accretion rate sources, but a more detailed explanation of this result would only be
possible by a deeper understanding of its role as a thermodynamic diagnostic tool.

\section{Discussion and conclusions}
The main purpose of this work is to collect a large sample of flux and profile measurements
of the broad emission line components in the Balmer series and to compare the results with
those AGN properties that we were able to infer from the available data. The paper is intended
as a starting point in a larger investigation aiming at the identifiaction of possible hints
about the physical conditions within the BLR of many AGN, through the analysis of various broad
lines.

While we spent a great effort in performing as carefully as possible our direct measurements on
the broad emission lines, the size of our sample and the limited spectral coverage forced us to
calculate the other source parameters by means of some simplified assumptions and empirical
relations, which, in some cases, are still matter of debate. Using the BP method, we explored
the relation of the temperature parameter $A$ and the broad line profiles. The values inferred
in those objects where the technique achieved the best results would correspond to a temperature
range running from $8\cdot 10^3\ {\rm K}$ to $21\cdot 10^3\ {\rm K}$, with cold gas being usually
associated with broad line emitters. A better explanation of this result, which could be of
fundamental importance in our understanding of the BLR structure, requires a deep investigation
of the properties of the temperature parameter $A$ as a thermodynamic tool \citep{ilic07}. 
The possibility of a relation with the inferred accretion rates, too, deserves further
investigation, since it may inform us about the different conditions of gas ionization or dust
processing of the radiation within our point of view towards the BLR.

Taking into account the results of our measurements and the limits of the related assumptions,
here we come to the following conclusions:
\begin{figure}[t]
\includegraphics[width = 8.2cm]{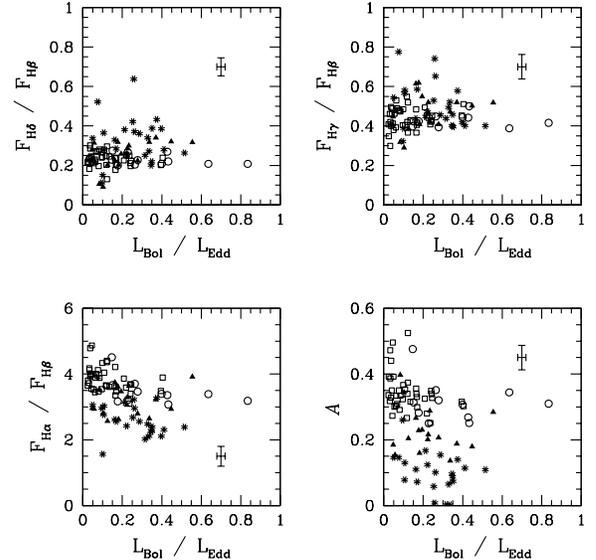}
\caption{Balmer line flux ratios of $H\delta$ (upper left panel), $H\gamma$ (upper right),
and $H\alpha$ (lower left) together with the BP slope parameter (lower right panel) as a
function of the Eddington ratio, with the same symbols as in Fig.~\ref{f09}. There appears
 to be a weak anti-correlation between the accretion rate and the Balmer decrement, with
the high order lines being stronger in objects working at high Eddington ratios, resulting
in modest BP slopes.\label{f11}}
\end{figure}
\begin{itemize}
\item the Balmer line flux ratios show a weak degree of correlation with the line profile width,
in the sense that more pronounced Balmer decrements are observed in broader line emitting 
objects;
\item the optical continuum luminosity of AGN is related to the emission line intensity of
[\ion{O}{3}], where we confirm the presence of a global Baldwin Effect, and of the broad
components of the Balmer series, which, instead, do not show any strong evidence of global
Baldwin Effect;
\item the shape of the Balmer decrement, studied with the Boltzmann Plot technique, is
related to the line profile widths and we observe that larger slopes and better straight line
fits can be achieved in objects with broader lines, where a BLR component being described by
the assumptions of this method may be present;
\item we find a weak degree of anti-correlation between the Balmer decrement and the Eddington
ratios measured in our sample, that could be the result of high dust column densities along our
line of sight to the objects characterized by low accretion rates.
\end{itemize}

\acknowledgments
We thank the anonymous referee for discussion and suggestions leading to the improvement
of this work.

GLM is particularly grateful to C. W. Yip for providing help and suggestions in the use
of eigenspectra. L. \v C. P. and D.I. were supported by the Ministry of Science and
Environment Protection of Serbia through the project 'Astrophysical Spectroscopy of
Extragalactic Objects'. Funding for the SDSS and SDSS-II has been provided by the Alfred P.
Sloan Foundation, the Participating Institutions, the National Science Foundation, the U.S.
Department of Energy, the National Aeronautics and Space Administration, the Japanese
Monbukagakusho, the Max Planck Society, and the Higher Education Funding Council for England.
The SDSS Web Site is http://www.sdss.org/.

The SDSS is managed by the Astrophysical Research Consortium for the Participating
Institutions. The Participating Institutions are the American Museum of Natural History,
Astrophysical Institute Potsdam, University of Basel, University of Cambridge, Case Western
Reserve University, University of Chicago, Drexel University, Fermilab, the Institute for
Advanced Study, the Japan Participation Group, Johns Hopkins University, the Joint Institute
for Nuclear Astrophysics, the Kavli Institute for Particle Astrophysics and Cosmology, the
Korean Scientist Group, the Chinese Academy of Sciences (LAMOST), Los Alamos National
Laboratory, the Max-Planck-Institute for Astronomy (MPIA), the Max-Planck-Institute for
Astrophysics (MPA), New Mexico State University, Ohio State University, University of
Pittsburgh, University of Portsmouth, Princeton University, the United States Naval
Observatory, and the University of Washington. This research has made use of NASA's
Astrophysics Data System.

\clearpage
\begin{deluxetable}{lcccccccc}
\tabletypesize{\tiny}
\tablecolumns{9}
\tablenum{1}
\tablecaption{Line Flux and Boltzmann Plot analysis\label{t01}}
\tablehead{\colhead{Object Name} & \colhead{$F_{\rm H\beta}$\tablenotemark{a}} & \colhead{$R_{\rm H\epsilon}$} & \colhead{$R_{\rm H\delta}$} &
\colhead{$R_{\rm H\gamma}$} & \colhead{$R_{\rm H\alpha}$} & \colhead{$F_{\rm [O\ III]}$\tablenotemark{a}} & \colhead{$A$} & \colhead{Class}}
\startdata
SDSSJ0013--0951 & 427 $\pm$ 54 & ND & 0.24 $\pm$ 0.05 & 0.51 $\pm$ 0.08 & 3.89 $\pm$ 0.54 & 872 $\pm$ 63 & 0.302 $\pm$ 0.033 & ii \\
SDSSJ0013+0052 & 1210 $\pm$ 108 & ND & 0.24 $\pm$ 0.04 & 0.42 $\pm$ 0.05 & 3.48 $\pm$ 0.33 & 237 $\pm$ 25 & 0.299 $\pm$ 0.011 & ii \\
SDSSJ0037+0008 & 1054 $\pm$ 95 & 0.09 $\pm$ 0.02 & 0.21 $\pm$ 0.03 & 0.39 $\pm$ 0.06 & 3.39 $\pm$ 0.36 & 632 $\pm$ 65 & 0.344 $\pm$ 0.038 & i \\
SDSSJ0042--1049 & 740 $\pm$ 87 & ND & 0.24 $\pm$ 0.05 & 0.49 $\pm$ 0.09 & 4.34 $\pm$ 0.56 & 637 $\pm$ 50 & 0.367 $\pm$ 0.032 & ii \\
SDSSJ0107+1408 & 954 $\pm$ 72 & 0.10 $\pm$ 0.03 & 0.21 $\pm$ 0.04 & 0.45 $\pm$ 0.06 & 3.40 $\pm$ 0.30 & 242 $\pm$ 22 & 0.307 $\pm$ 0.023 & i \\
SDSSJ0110--1008 & 5856 $\pm$ 452 & 0.10 $\pm$ 0.02 & 0.21 $\pm$ 0.03 & 0.43 $\pm$ 0.05 & 3.69 $\pm$ 0.31 & 3154 $\pm$ 126 & 0.349 $\pm$ 0.016 & ii \\
SDSSJ0112+0003 & 702 $\pm$ 82 & ND & 0.28 $\pm$ 0.06 & 0.40 $\pm$ 0.08 & 4.04 $\pm$ 0.55 & 611 $\pm$ 16 & 0.353 $\pm$ 0.053 & ii \\
SDSSJ0117+0000 & 956 $\pm$ 40 & 0.13 $\pm$ 0.04 & 0.27 $\pm$ 0.04 & 0.44 $\pm$ 0.03 & 3.22 $\pm$ 0.17 & 1194 $\pm$ 35 & 0.250 $\pm$ 0.015 & i \\
SDSSJ0121--0102 & 1860 $\pm$ 137 & ND & 0.26 $\pm$ 0.04 & 0.40 $\pm$ 0.05 & 2.38 $\pm$ 0.23 & 334 $\pm$ 50 & 0.109 $\pm$ 0.049 & iv \\
SDSSJ0135--0044 & 4053 $\pm$ 227 & 0.06 $\pm$ 0.01 & 0.20 $\pm$ 0.03 & 0.42 $\pm$ 0.04 & 3.40 $\pm$ 0.23 & 771 $\pm$ 60 & 0.339 $\pm$ 0.061 & ii \\
SDSSJ0140--0050 & 1315 $\pm$ 100 & ND & 0.20 $\pm$ 0.05 & 0.41 $\pm$ 0.06 & 2.62 $\pm$ 0.25 & 473 $\pm$ 45 & 0.160 $\pm$ 0.044 & iii \\
SDSSJ0142--1008 & 16431 $\pm$ 788 & ND & 0.22 $\pm$ 0.03 & 0.53 $\pm$ 0.06 & 3.89 $\pm$ 0.22 & 14866 $\pm$ 536 & 0.334 $\pm$ 0.048 & ii \\
SDSSJ0142+0005 & 345 $\pm$ 37 & ND & 0.37 $\pm$ 0.09 & 0.65 $\pm$ 0.13 & 2.96 $\pm$ 0.37 & 233 $\pm$ 13 & 0.101 $\pm$ 0.071 & iv \\
SDSSJ0150+1323 & 4297 $\pm$ 379 & ND & 0.18 $\pm$ 0.03 & 0.30 $\pm$ 0.05 & 4.18 $\pm$ 0.41 & 2272 $\pm$ 80 & 0.472 $\pm$ 0.041 & ii \\
SDSSJ0159+0105 & 935 $\pm$ 100 & ND & 0.21 $\pm$ 0.05 & 0.41 $\pm$ 0.07 & 3.60 $\pm$ 0.42 & 296 $\pm$ 22 & 0.330 $\pm$ 0.008 & ii \\
SDSSJ0233--0107 & 1067 $\pm$ 87 & ND & 0.20 $\pm$ 0.04 & 0.42 $\pm$ 0.05 & 3.63 $\pm$ 0.34 & 663 $\pm$ 33 & 0.340 $\pm$ 0.012 & ii \\
SDSSJ0250+0025 & 940 $\pm$ 65 & ND & 0.32 $\pm$ 0.06 & 0.62 $\pm$ 0.08 & 3.62 $\pm$ 0.28 & 613 $\pm$ 37 & 0.229 $\pm$ 0.068 & iii \\
SDSSJ0256+0113 & 964 $\pm$ 82 & ND & 0.21 $\pm$ 0.04 & 0.46 $\pm$ 0.06 & 3.54 $\pm$ 0.32 & 129 $\pm$ 21 & 0.301 $\pm$ 0.017 & ii \\
SDSSJ0304+0028 & 618 $\pm$ 67 & ND & 0.27 $\pm$ 0.06 & 0.52 $\pm$ 0.09 & 2.11 $\pm$ 0.27 & 174 $\pm$ 24 & 0.002 $\pm$ 0.024 & iv \\
SDSSJ0306+0003 & 496 $\pm$ 94 & ND & 0.38 $\pm$ 0.11 & 0.45 $\pm$ 0.11 & 3.14 $\pm$ 0.63 & 138 $\pm$ 8 & 0.167 $\pm$ 0.071 & iv \\
SDSSJ0310--0049 & 1953 $\pm$ 113 & ND & 0.09 $\pm$ 0.02 & 0.29 $\pm$ 0.03 & 2.81 $\pm$ 0.21 & 293 $\pm$ 33 & 0.344 $\pm$ 0.137 & iii \\
SDSSJ0322+0055 & 564 $\pm$ 68 & ND & 0.24 $\pm$ 0.05 & 0.48 $\pm$ 0.08 & 4.22 $\pm$ 0.56 & 408 $\pm$ 50 & 0.356 $\pm$ 0.031 & ii \\
SDSSJ0323+0035 & 5994 $\pm$ 288 & 0.08 $\pm$ 0.01 & 0.21 $\pm$ 0.02 & 0.42 $\pm$ 0.04 & 3.19 $\pm$ 0.18 & 1729 $\pm$ 222 & 0.310 $\pm$ 0.046 & i \\
SDSSJ0351--0526 & 7617 $\pm$ 454 & 0.12 $\pm$ 0.02 & 0.25 $\pm$ 0.03 & 0.47 $\pm$ 0.06 & 2.42 $\pm$ 0.18 & 6649 $\pm$ 310 & 0.124 $\pm$ 0.040 & iv \\
SDSSJ0409--0429 & 4767 $\pm$ 310 & ND & 0.18 $\pm$ 0.03 & 0.37 $\pm$ 0.05 & 3.35 $\pm$ 0.26 & 1385 $\pm$ 42 & 0.339 $\pm$ 0.035 & ii \\
SDSSJ0752+2617 & 487 $\pm$ 84 & ND & 0.33 $\pm$ 0.10 & 0.44 $\pm$ 0.12 & 2.61 $\pm$ 0.50 & 295 $\pm$ 24 & 0.110 $\pm$ 0.045 & iv \\
SDSSJ0755+3911 & 2868 $\pm$ 290 & 0.14 $\pm$ 0.03 & 0.42 $\pm$ 0.07 & 0.53 $\pm$ 0.08 & 2.68 $\pm$ 0.30 & 1914 $\pm$ 63 & 0.058 $\pm$ 0.061 & iv \\
SDSSJ0830+3405 & 4102 $\pm$ 265 & ND & 0.25 $\pm$ 0.04 & 0.46 $\pm$ 0.07 & 4.00 $\pm$ 0.29 & 1582 $\pm$ 92 & 0.350 $\pm$ 0.029 & ii \\
SDSSJ0832+4614 & 2902 $\pm$ 217 & 0.10 $\pm$ 0.02 & 0.27 $\pm$ 0.04 & 0.42 $\pm$ 0.05 & 3.66 $\pm$ 0.30 & 1388 $\pm$ 54 & 0.314 $\pm$ 0.038 & i \\
SDSSJ0839+4847 & 10536 $\pm$ 588 & ND & 0.21 $\pm$ 0.04 & 0.41 $\pm$ 0.05 & 3.65 $\pm$ 0.21 & 1164 $\pm$ 41 & 0.336 $\pm$ 0.002 & ii \\
SDSSJ0840+0333 & 938 $\pm$ 73 & ND & 0.23 $\pm$ 0.04 & 0.46 $\pm$ 0.06 & 3.72 $\pm$ 0.32 & 439 $\pm$ 19 & 0.318 $\pm$ 0.018 & ii \\
SDSSJ0855+5252 & 991 $\pm$ 122 & ND & 0.23 $\pm$ 0.05 & 0.49 $\pm$ 0.10 & 3.86 $\pm$ 0.51 & 430 $\pm$ 42 & 0.325 $\pm$ 0.032 & ii \\
SDSSJ0857+0528 & 959 $\pm$ 127 & ND & 0.52 $\pm$ 0.13 & 0.77 $\pm$ 0.17 & 3.90 $\pm$ 0.55 & 1127 $\pm$ 52 & 0.146 $\pm$ 0.122 & iv \\
SDSSJ0904+5536 & 699 $\pm$ 120 & ND & 0.24 $\pm$ 0.06 & 0.49 $\pm$ 0.12 & 3.55 $\pm$ 0.63 & 862 $\pm$ 24 & 0.273 $\pm$ 0.019 & ii \\
SDSSJ0925+5335 & 406 $\pm$ 50 & ND & 0.64 $\pm$ 0.15 & 0.74 $\pm$ 0.14 & 3.22 $\pm$ 0.48 & 416 $\pm$ 22 & 0.008 $\pm$ 0.142 & iv \\
SDSSJ0937+0105 & 1684 $\pm$ 92 & ND & 0.26 $\pm$ 0.04 & 0.55 $\pm$ 0.05 & 3.47 $\pm$ 0.22 & 1352 $\pm$ 76 & 0.230 $\pm$ 0.048 & iii \\
SDSSJ1010+0043 & 2334 $\pm$ 139 & 0.11 $\pm$ 0.00 & 0.20 $\pm$ 0.02 & 0.45 $\pm$ 0.04 & 3.70 $\pm$ 0.28 & 1731 $\pm$ 64 & 0.351 $\pm$ 0.026 & i \\
SDSSJ1013--0052 & 2300 $\pm$ 114 & ND & 0.15 $\pm$ 0.03 & 0.39 $\pm$ 0.05 & 1.56 $\pm$ 0.09 & 417 $\pm$ 54 & -0.109 $\pm$ 0.117 & iv \\
SDSSJ1016+4210 & 5835 $\pm$ 497 & 0.10 $\pm$ 0.02 & 0.36 $\pm$ 0.06 & 0.46 $\pm$ 0.07 & 2.56 $\pm$ 0.28 & 1497 $\pm$ 198 & 0.154 $\pm$ 0.096 & iv \\
SDSSJ1025+5140 & 2045 $\pm$ 133 & ND & 0.30 $\pm$ 0.04 & 0.42 $\pm$ 0.05 & 3.64 $\pm$ 0.30 & 2349 $\pm$ 68 & 0.288 $\pm$ 0.057 & iii \\
SDSSJ1042+0414 & 755 $\pm$ 71 & ND & 0.22 $\pm$ 0.05 & 0.61 $\pm$ 0.09 & 3.75 $\pm$ 0.40 & 1037 $\pm$ 66 & 0.266 $\pm$ 0.076 & iii \\
SDSSJ1057--0041 & 1087 $\pm$ 76 & ND & 0.37 $\pm$ 0.07 & 0.57 $\pm$ 0.04 & 2.76 $\pm$ 0.23 & 357 $\pm$ 29 & 0.078 $\pm$ 0.035 & iv \\
SDSSJ1059--0005 & 934 $\pm$ 86 & ND & 0.22 $\pm$ 0.04 & 0.46 $\pm$ 0.09 & 2.26 $\pm$ 0.23 & 162 $\pm$ 32 & 0.087 $\pm$ 0.046 & iv \\
SDSSJ1105+0745 & 1610 $\pm$ 186 & ND & 0.22 $\pm$ 0.06 & 0.35 $\pm$ 0.06 & 3.80 $\pm$ 0.47 & 589 $\pm$ 25 & 0.392 $\pm$ 0.032 & ii \\
SDSSJ1118+5803 & 10103 $\pm$ 594 & ND & 0.27 $\pm$ 0.04 & 0.48 $\pm$ 0.06 & 4.12 $\pm$ 0.28 & 7199 $\pm$ 279 & 0.350 $\pm$ 0.047 & ii \\
SDSSJ1122+0117 & 1791 $\pm$ 115 & ND & 0.22 $\pm$ 0.04 & 0.45 $\pm$ 0.06 & 3.68 $\pm$ 0.32 & 2470 $\pm$ 69 & 0.327 $\pm$ 0.017 & ii \\
SDSSJ1128+1023 & 2622 $\pm$ 262 & ND & 0.32 $\pm$ 0.08 & 0.52 $\pm$ 0.08 & 3.92 $\pm$ 0.45 & 2644 $\pm$ 208 & 0.284 $\pm$ 0.059 & iii \\
SDSSJ1139+5911 & 1654 $\pm$ 160 & 0.16 $\pm$ 0.05 & 0.28 $\pm$ 0.06 & 0.59 $\pm$ 0.09 & 2.56 $\pm$ 0.31 & 887 $\pm$ 52 & 0.072 $\pm$ 0.031 & iv \\
SDSSJ1141+0241 & 1109 $\pm$ 285 & ND & 0.28 $\pm$ 0.11 & 0.58 $\pm$ 0.18 & 3.03 $\pm$ 0.80 & 377 $\pm$ 13 & 0.130 $\pm$ 0.031 & iv \\
SDSSJ1152--0005 & 859 $\pm$ 79 & 0.20 $\pm$ 0.06 & 0.25 $\pm$ 0.05 & 0.44 $\pm$ 0.08 & 2.02 $\pm$ 0.20 & 332 $\pm$ 48 & 0.003 $\pm$ 0.031 & iv \\
SDSSJ1157--0022 & 5477 $\pm$ 177 & ND & 0.11 $\pm$ 0.02 & 0.32 $\pm$ 0.04 & 2.96 $\pm$ 0.13 & 990 $\pm$ 48 & 0.260 $\pm$ 0.084 & iv \\
SDSSJ1157+0412 & 2188 $\pm$ 195 & 0.13 $\pm$ 0.02 & 0.27 $\pm$ 0.06 & 0.44 $\pm$ 0.08 & 3.36 $\pm$ 0.31 & 739 $\pm$ 111 & 0.268 $\pm$ 0.014 & i \\
SDSSJ1203+0229 & 3719 $\pm$ 270 & 0.12 $\pm$ 0.02 & 0.20 $\pm$ 0.03 & 0.42 $\pm$ 0.04 & 3.17 $\pm$ 0.24 & 501 $\pm$ 80 & 0.282 $\pm$ 0.016 & i \\
SDSSJ1223+0240 & 870 $\pm$ 97 & ND & 0.25 $\pm$ 0.06 & 0.55 $\pm$ 0.09 & 3.66 $\pm$ 0.43 & 987 $\pm$ 40 & 0.266 $\pm$ 0.039 & ii \\
SDSSJ1243+0252 & 738 $\pm$ 51 & ND & 0.20 $\pm$ 0.03 & 0.45 $\pm$ 0.06 & 3.49 $\pm$ 0.28 & 132 $\pm$ 24 & 0.316 $\pm$ 0.023 & ii \\
SDSSJ1246+0222 & 3619 $\pm$ 160 & ND & 0.29 $\pm$ 0.03 & 0.52 $\pm$ 0.04 & 3.29 $\pm$ 0.17 & 1982 $\pm$ 73 & 0.216 $\pm$ 0.036 & iii \\
SDSSJ1300+5641 & 1731 $\pm$ 146 & ND & 0.25 $\pm$ 0.06 & 0.45 $\pm$ 0.06 & 3.59 $\pm$ 0.32 & 1088 $\pm$ 43 & 0.280 $\pm$ 0.019 & ii \\
SDSSJ1300+6139 & 2488 $\pm$ 128 & ND & 0.30 $\pm$ 0.05 & 0.50 $\pm$ 0.06 & 3.93 $\pm$ 0.27 & 1105 $\pm$ 54 & 0.328 $\pm$ 0.061 & ii \\
SDSSJ1307--0036 & 1204 $\pm$ 80 & 0.08 $\pm$ 0.02 & 0.22 $\pm$ 0.03 & 0.41 $\pm$ 0.05 & 2.79 $\pm$ 0.20 & 537 $\pm$ 35 & 0.207 $\pm$ 0.039 & iii \\
SDSSJ1307+0107 & 1956 $\pm$ 109 & 0.06 $\pm$ 0.01 & 0.23 $\pm$ 0.04 & 0.40 $\pm$ 0.04 & 4.51 $\pm$ 0.40 & 1389 $\pm$ 56 & 0.476 $\pm$ 0.075 & i \\
SDSSJ1331+0131 & 2753 $\pm$ 268 & 0.17 $\pm$ 0.03 & 0.39 $\pm$ 0.08 & 0.40 $\pm$ 0.06 & 2.40 $\pm$ 0.26 & 2120 $\pm$ 269 & 0.074 $\pm$ 0.059 & iv \\
SDSSJ1341--0053 & 1433 $\pm$ 99 & 0.11 $\pm$ 0.02 & 0.23 $\pm$ 0.03 & 0.39 $\pm$ 0.05 & 3.46 $\pm$ 0.30 & 1174 $\pm$ 42 & 0.320 $\pm$ 0.020 & i \\
SDSSJ1342--0053 & 840 $\pm$ 90 & ND & 0.27 $\pm$ 0.05 & 0.49 $\pm$ 0.07 & 4.00 $\pm$ 0.47 & 355 $\pm$ 25 & 0.309 $\pm$ 0.038 & ii \\
SDSSJ1342+5642 & 903 $\pm$ 78 & ND & 0.20 $\pm$ 0.05 & 0.48 $\pm$ 0.08 & 3.47 $\pm$ 0.34 & 155 $\pm$ 13 & 0.291 $\pm$ 0.030 & ii \\
SDSSJ1343+0004 & 1269 $\pm$ 122 & 0.14 $\pm$ 0.03 & 0.34 $\pm$ 0.06 & 0.52 $\pm$ 0.07 & 3.23 $\pm$ 0.36 & 462 $\pm$ 53 & 0.188 $\pm$ 0.043 & iii \\
SDSSJ1344--0015 & 1577 $\pm$ 136 & 0.14 $\pm$ 0.03 & 0.34 $\pm$ 0.05 & 0.49 $\pm$ 0.05 & 2.48 $\pm$ 0.84 & 406 $\pm$ 117 & 0.065 $\pm$ 0.109 & iv \\
SDSSJ1344+0005 & 1912 $\pm$ 103 & ND & 0.11 $\pm$ 0.02 & 0.32 $\pm$ 0.04 & 3.53 $\pm$ 0.22 & 428 $\pm$ 31 & 0.397 $\pm$ 0.073 & iii \\
SDSSJ1344+4416 & 2292 $\pm$ 238 & ND & 0.39 $\pm$ 0.08 & 0.42 $\pm$ 0.08 & 2.11 $\pm$ 0.24 & 512 $\pm$ 81 & -0.017 $\pm$ 0.062 & iv \\
SDSSJ1345--0259 & 3518 $\pm$ 243 & ND & 0.31 $\pm$ 0.04 & 0.47 $\pm$ 0.05 & 2.94 $\pm$ 0.23 & 1317 $\pm$ 46 & 0.154 $\pm$ 0.033 & iii \\
SDSSJ1349+0204 & 3930 $\pm$ 220 & ND & 0.22 $\pm$ 0.03 & 0.41 $\pm$ 0.06 & 4.87 $\pm$ 0.30 & 8162 $\pm$ 260 & 0.496 $\pm$ 0.030 & ii \\
SDSSJ1355+6440 & 4383 $\pm$ 352 & 0.13 $\pm$ 0.02 & 0.32 $\pm$ 0.05 & 0.47 $\pm$ 0.06 & 2.65 $\pm$ 0.23 & 2483 $\pm$ 163 & 0.136 $\pm$ 0.033 & iii \\
SDSSJ1437+0007 & 1321 $\pm$ 87 & 0.12 $\pm$ 0.02 & 0.22 $\pm$ 0.02 & 0.50 $\pm$ 0.06 & 3.07 $\pm$ 0.25 & 335 $\pm$ 43 & 0.251 $\pm$ 0.029 & i \\
SDSSJ1505+0342 & 1348 $\pm$ 144 & 0.14 $\pm$ 0.04 & 0.43 $\pm$ 0.08 & 0.58 $\pm$ 0.10 & 3.27 $\pm$ 0.38 & 1202 $\pm$ 29 & 0.140 $\pm$ 0.066 & iv \\
\enddata
\end{deluxetable}
\begin{deluxetable}{lcccccccc}
\tabletypesize{\tiny}
\tablecolumns{9}
\tablenum{1}
\tablecaption{Line Flux and Boltzmann Plot analysis}
\tablehead{\colhead{Object Name} & \colhead{$F_{\rm H\beta}$\tablenotemark{a}} & \colhead{$R_{\rm H\epsilon}$} & \colhead{$R_{\rm H\delta}$} &
\colhead{$R_{\rm H\gamma}$} & \colhead{$R_{\rm H\alpha}$} & \colhead{$F_{\rm [O\ III]}$\tablenotemark{a}} & \colhead{$A$} & \colhead{Class}}
\startdata
SDSSJ1510+0058 & 10659 $\pm$ 695 & ND & 0.23 $\pm$ 0.03 & 0.38 $\pm$ 0.04 & 4.00 $\pm$ 0.31 & 15250 $\pm$ 377 & 0.392 $\pm$ 0.027 & ii \\
SDSSJ1519+0016 & 1306 $\pm$ 133 & ND & 0.25 $\pm$ 0.06 & 0.43 $\pm$ 0.05 & 3.61 $\pm$ 0.45 & 392 $\pm$ 27 & 0.316 $\pm$ 0.013 & ii \\
SDSSJ1519+5908 & 1596 $\pm$ 131 & 0.13 $\pm$ 0.03 & 0.32 $\pm$ 0.07 & 0.50 $\pm$ 0.07 & 2.94 $\pm$ 0.26 & 416 $\pm$ 72 & 0.179 $\pm$ 0.034 & iii \\
SDSSJ1535+5754 & 1658 $\pm$ 122 & ND & 0.23 $\pm$ 0.04 & 0.39 $\pm$ 0.05 & 3.45 $\pm$ 0.28 & 238 $\pm$ 25 & 0.304 $\pm$ 0.018 & ii \\
SDSSJ1538+4440 & 1212 $\pm$ 90 & ND & 0.28 $\pm$ 0.05 & 0.45 $\pm$ 0.07 & 2.96 $\pm$ 0.25 & 1017 $\pm$ 30 & 0.185 $\pm$ 0.020 & iii \\
SDSSJ1554+3238 & 3053 $\pm$ 283 & ND & 0.29 $\pm$ 0.06 & 0.40 $\pm$ 0.07 & 4.78 $\pm$ 0.46 & 3065 $\pm$ 116 & 0.440 $\pm$ 0.061 & ii \\
SDSSJ1613+3717 & 1797 $\pm$ 193 & ND & 0.24 $\pm$ 0.05 & 0.42 $\pm$ 0.06 & 4.12 $\pm$ 0.48 & 528 $\pm$ 17 & 0.387 $\pm$ 0.022 & ii \\
SDSSJ1619+4058 & 2609 $\pm$ 164 & ND & 0.30 $\pm$ 0.05 & 0.52 $\pm$ 0.08 & 4.40 $\pm$ 0.31 & 6830 $\pm$ 293 & 0.374 $\pm$ 0.070 & ii \\
SDSSJ1623+4104 & 6033 $\pm$ 481 & 0.09 $\pm$ 0.02 & 0.28 $\pm$ 0.05 & 0.40 $\pm$ 0.06 & 2.31 $\pm$ 0.20 & 1254 $\pm$ 186 & 0.114 $\pm$ 0.056 & iv \\
SDSSJ1654+3925 & 2396 $\pm$ 178 & ND & 0.28 $\pm$ 0.07 & 0.47 $\pm$ 0.07 & 3.88 $\pm$ 0.32 & 3564 $\pm$ 131 & 0.334 $\pm$ 0.037 & ii \\
SDSSJ1659+6202 & 1440 $\pm$ 101 & ND & 0.25 $\pm$ 0.04 & 0.41 $\pm$ 0.05 & 3.10 $\pm$ 0.24 & 149 $\pm$ 27 & 0.251 $\pm$ 0.019 & ii \\
SDSSJ1717+5815 & 1906 $\pm$ 144 & ND & 0.20 $\pm$ 0.04 & 0.40 $\pm$ 0.07 & 2.27 $\pm$ 0.22 & 521 $\pm$ 49 & 0.096 $\pm$ 0.065 & iv \\
SDSSJ1719+5937 & 5646 $\pm$ 297 & ND & 0.13 $\pm$ 0.02 & 0.37 $\pm$ 0.04 & 4.36 $\pm$ 0.34 & 1955 $\pm$ 144 & 0.525 $\pm$ 0.072 & ii \\
SDSSJ1720+5540 & 3722 $\pm$ 253 & ND & 0.24 $\pm$ 0.04 & 0.40 $\pm$ 0.04 & 3.54 $\pm$ 0.26 & 5491 $\pm$ 213 & 0.323 $\pm$ 0.016 & ii \\
SDSSJ2058--0650 & 1116 $\pm$ 191 & ND & 0.20 $\pm$ 0.06 & 0.39 $\pm$ 0.10 & 2.57 $\pm$ 0.48 & 296 $\pm$ 30 & 0.204 $\pm$ 0.036 & iii \\
SDSSJ2349--0036 & 1837 $\pm$ 176 & ND & 0.34 $\pm$ 0.06 & 0.54 $\pm$ 0.08 & 3.05 $\pm$ 0.33 & 519 $\pm$ 45 & 0.146 $\pm$ 0.043 & iv \\
SDSSJ2351--0109 & 522 $\pm$ 52 & ND & 0.26 $\pm$ 0.04 & 0.47 $\pm$ 0.09 & 3.06 $\pm$ 0.35 & 186 $\pm$ 17 & 0.201 $\pm$ 0.013 & iii \\
\enddata
\tablenotetext{a}{Fluxes are given in units of $10^{-17} {\rm erg\ cm^{-2}\ s^{-1}}$}
\end{deluxetable}

\begin{deluxetable}{lcccccc}
\tabletypesize{\tiny}
\tablecolumns{7}
\tablenum{2}
\tablecaption{Line profile and source property estimates\label{t02}}
\tablehead{\colhead{Object Name} & \colhead{FWHM\tablenotemark{a}} & \colhead{FWZI\tablenotemark{a}} &
\colhead{$L_{5100}$\tablenotemark{b}} & \colhead{$R_{\rm BLR}$\tablenotemark{c}} &
\colhead{$M_{\rm BH}$\tablenotemark{d}} & \colhead{$\dot{M} / \dot{M}_{\rm Edd}$}}
\startdata
SDSSJ0013--0951 & 1122 $\pm$ 150 & 5951 $\pm$ 1632 & 3.86 $\pm$ 0.17 & 11.35 $\pm$ 1.31 & 8.08 $\pm$ 0.93 & 0.405 $\pm$ 0.064 \\
SDSSJ0013+0052 & 4163 $\pm$ 234 & 10573 $\pm$ 1458 & 128.04 $\pm$ 1.06 & 65.35 $\pm$ 6.43 & 639.96 $\pm$ 62.92 & 0.170 $\pm$ 0.018 \\
SDSSJ0037+0008 & 2608 $\pm$ 150 & 8744 $\pm$ 1467 & 277.83 $\pm$ 4.94 & 96.27 $\pm$ 9.92 & 369.99 $\pm$ 38.13 & 0.636 $\pm$ 0.077 \\
SDSSJ0042--1049 & 2272 $\pm$ 150 & 7478 $\pm$ 1355 & 3.89 $\pm$ 0.08 & 11.40 $\pm$ 1.19 & 33.23 $\pm$ 3.46 & 0.099 $\pm$ 0.012 \\
SDSSJ0107+1408 & 2214 $\pm$ 150 & 8956 $\pm$ 1661 & 56.76 $\pm$ 1.39 & 43.51 $\pm$ 4.63 & 120.52 $\pm$ 12.83 & 0.399 $\pm$ 0.052 \\
SDSSJ0110--1008 & 2808 $\pm$ 216 & 10537 $\pm$ 1261 & 53.19 $\pm$ 0.46 & 42.12 $\pm$ 4.15 & 187.64 $\pm$ 18.49 & 0.240 $\pm$ 0.026 \\
SDSSJ0112+0003 & 2218 $\pm$ 179 & 8524 $\pm$ 2044 & 4.16 $\pm$ 0.05 & 11.78 $\pm$ 1.19 & 32.75 $\pm$ 3.29 & 0.108 $\pm$ 0.012 \\
SDSSJ0117+0000 & 2571 $\pm$ 243 & 8610 $\pm$ 1831 & 33.58 $\pm$ 1.53 & 33.47 $\pm$ 3.91 & 124.96 $\pm$ 14.61 & 0.228 $\pm$ 0.037 \\
SDSSJ0121--0102 & 3253 $\pm$ 150 & 9969 $\pm$ 1480 & 439.86 $\pm$ 2.73 & 121.13 $\pm$ 11.78 & 724.50 $\pm$ 70.48 & 0.515 $\pm$ 0.053 \\
SDSSJ0135--0044 & 5586 $\pm$ 379 & 12992 $\pm$ 1813 & 845.06 $\pm$ 14.18 & 167.90 $\pm$ 17.22 & 2960.01 $\pm$ 303.58 & 0.242 $\pm$ 0.029 \\
SDSSJ0140--0050 & 2621 $\pm$ 150 & 9068 $\pm$ 1374 & 22.95 $\pm$ 0.33 & 27.67 $\pm$ 2.80 & 107.38 $\pm$ 10.88 & 0.181 $\pm$ 0.021 \\
SDSSJ0142--1008 & 4266 $\pm$ 150 & 12344 $\pm$ 2467 & 20.11 $\pm$ 0.31 & 25.90 $\pm$ 2.64 & 266.31 $\pm$ 27.10 & 0.064 $\pm$ 0.007 \\
SDSSJ0142+0005 & 1142 $\pm$ 150 & 5568 $\pm$ 1203 & 1.73 $\pm$ 0.08 & 7.59 $\pm$ 0.88 & 5.59 $\pm$ 0.65 & 0.262 $\pm$ 0.042 \\
SDSSJ0150+1323 & 4846 $\pm$ 410 & 14046 $\pm$ 3127 & 8.78 $\pm$ 0.13 & 17.12 $\pm$ 1.73 & 227.13 $\pm$ 23.01 & 0.033 $\pm$ 0.004 \\
SDSSJ0159+0105 & 3077 $\pm$ 201 & 9978 $\pm$ 2151 & 37.36 $\pm$ 0.65 & 35.30 $\pm$ 3.63 & 188.85 $\pm$ 19.42 & 0.168 $\pm$ 0.020 \\
SDSSJ0233--0107 & 3330 $\pm$ 291 & 9277 $\pm$ 1278 & 26.13 $\pm$ 0.50 & 29.52 $\pm$ 3.07 & 184.98 $\pm$ 19.20 & 0.120 $\pm$ 0.015 \\
SDSSJ0250+0025 & 1453 $\pm$ 150 & 6686 $\pm$ 1400 & 2.07 $\pm$ 0.07 & 8.32 $\pm$ 0.91 & 9.93 $\pm$ 1.09 & 0.177 $\pm$ 0.025 \\
SDSSJ0256+0113 & 3504 $\pm$ 150 & 8969 $\pm$ 1945 & 6.76 $\pm$ 0.18 & 15.01 $\pm$ 1.61 & 104.17 $\pm$ 11.19 & 0.055 $\pm$ 0.007 \\
SDSSJ0304+0028 & 2922 $\pm$ 157 & 8149 $\pm$ 1091 & 120.96 $\pm$ 1.85 & 63.52 $\pm$ 6.47 & 306.39 $\pm$ 31.19 & 0.335 $\pm$ 0.039 \\
SDSSJ0306+0003 & 1474 $\pm$ 185 & 7004 $\pm$ 1460 & 3.14 $\pm$ 0.13 & 10.23 $\pm$ 1.18 & 12.56 $\pm$ 1.44 & 0.212 $\pm$ 0.033 \\
SDSSJ0310--0049 & 4372 $\pm$ 150 & 10846 $\pm$ 1882 & 57.01 $\pm$ 1.14 & 43.61 $\pm$ 4.54 & 471.05 $\pm$ 49.07 & 0.103 $\pm$ 0.013 \\
SDSSJ0322+0055 & 2005 $\pm$ 150 & 6805 $\pm$ 1110 & 6.44 $\pm$ 0.22 & 14.66 $\pm$ 1.64 & 33.31 $\pm$ 3.72 & 0.164 $\pm$ 0.024 \\
SDSSJ0323+0035 & 2356 $\pm$ 150 & 9902 $\pm$ 1747 & 318.83 $\pm$ 2.38 & 103.13 $\pm$ 10.10 & 323.42 $\pm$ 31.66 & 0.835 $\pm$ 0.088 \\
SDSSJ0351--0526 & 2563 $\pm$ 162 & 8176 $\pm$ 1738 & 31.62 $\pm$ 0.72 & 32.48 $\pm$ 3.43 & 120.58 $\pm$ 12.73 & 0.222 $\pm$ 0.029 \\
SDSSJ0409--0429 & 3181 $\pm$ 154 & 9837 $\pm$ 1301 & 41.92 $\pm$ 0.42 & 37.40 $\pm$ 3.71 & 213.82 $\pm$ 21.21 & 0.166 $\pm$ 0.018 \\
SDSSJ0752+2617 & 2009 $\pm$ 150 & 7978 $\pm$ 1915 & 6.39 $\pm$ 0.17 & 14.60 $\pm$ 1.57 & 33.30 $\pm$ 3.58 & 0.163 $\pm$ 0.022 \\
SDSSJ0755+3911 & 1697 $\pm$ 150 & 7686 $\pm$ 1003 & 7.88 $\pm$ 0.09 & 16.21 $\pm$ 1.62 & 26.39 $\pm$ 2.63 & 0.253 $\pm$ 0.028 \\
SDSSJ0830+3405 & 5460 $\pm$ 239 & 13492 $\pm$ 1879 & 21.87 $\pm$ 0.55 & 27.01 $\pm$ 2.89 & 454.94 $\pm$ 48.60 & 0.041 $\pm$ 0.005 \\
SDSSJ0832+4614 & 2199 $\pm$ 176 & 10423 $\pm$ 1518 & 7.87 $\pm$ 0.16 & 16.21 $\pm$ 1.69 & 44.30 $\pm$ 4.62 & 0.151 $\pm$ 0.019 \\
SDSSJ0839+4847 & 5526 $\pm$ 179 & 14898 $\pm$ 3208 & 7.86 $\pm$ 0.13 & 16.19 $\pm$ 1.66 & 279.34 $\pm$ 28.70 & 0.024 $\pm$ 0.003 \\
SDSSJ0840+0333 & 4227 $\pm$ 234 & 9804 $\pm$ 1742 & 4.57 $\pm$ 0.06 & 12.35 $\pm$ 1.24 & 124.70 $\pm$ 12.51 & 0.031 $\pm$ 0.003 \\
SDSSJ0855+5252 & 1736 $\pm$ 150 & 6162 $\pm$ 1113 & 5.81 $\pm$ 0.23 & 13.92 $\pm$ 1.58 & 23.72 $\pm$ 2.69 & 0.208 $\pm$ 0.032 \\
SDSSJ0857+0528 & 2359 $\pm$ 150 & 7423 $\pm$ 1429 & 2.64 $\pm$ 0.02 & 9.39 $\pm$ 0.93 & 29.51 $\pm$ 2.91 & 0.076 $\pm$ 0.008 \\
SDSSJ0904+5536 & 2483 $\pm$ 163 & 6940 $\pm$ 1361 & 1.09 $\pm$ 0.01 & 6.02 $\pm$ 0.60 & 20.98 $\pm$ 2.10 & 0.044 $\pm$ 0.005 \\
SDSSJ0925+5335 & 1532 $\pm$ 150 & 5811 $\pm$ 1291 & 5.44 $\pm$ 0.07 & 13.47 $\pm$ 1.35 & 17.86 $\pm$ 1.79 & 0.258 $\pm$ 0.029 \\
SDSSJ0937+0105 & 2258 $\pm$ 150 & 7719 $\pm$ 1285 & 14.45 $\pm$ 0.53 & 21.95 $\pm$ 2.47 & 63.25 $\pm$ 7.12 & 0.194 $\pm$ 0.029 \\
SDSSJ1010+0043 & 2806 $\pm$ 150 & 8138 $\pm$ 1723 & 64.08 $\pm$ 2.98 & 46.23 $\pm$ 5.43 & 205.76 $\pm$ 24.16 & 0.264 $\pm$ 0.043 \\
SDSSJ1013--0052 & 6816 $\pm$ 164 & 11309 $\pm$ 1542 & 324.69 $\pm$ 3.59 & 104.07 $\pm$ 10.38 & 2732.11 $\pm$ 272.37 & 0.101 $\pm$ 0.011 \\
SDSSJ1016+4210 & 1845 $\pm$ 150 & 9904 $\pm$ 2052 & 14.11 $\pm$ 0.43 & 21.70 $\pm$ 2.37 & 41.73 $\pm$ 4.56 & 0.287 $\pm$ 0.040 \\
SDSSJ1025+5140 & 1845 $\pm$ 226 & 9883 $\pm$ 1104 & 9.42 $\pm$ 0.08 & 17.73 $\pm$ 1.75 & 34.11 $\pm$ 3.36 & 0.234 $\pm$ 0.025 \\
SDSSJ1042+0414 & 2082 $\pm$ 150 & 7751 $\pm$ 1499 & 7.41 $\pm$ 0.17 & 15.72 $\pm$ 1.67 & 38.49 $\pm$ 4.08 & 0.163 $\pm$ 0.021 \\
SDSSJ1057--0041 & 2370 $\pm$ 150 & 8870 $\pm$ 2027 & 5.28 $\pm$ 0.13 & 13.27 $\pm$ 1.41 & 42.13 $\pm$ 4.47 & 0.106 $\pm$ 0.014 \\
SDSSJ1059--0005 & 2674 $\pm$ 174 & 9662 $\pm$ 1278 & 92.33 $\pm$ 1.12 & 55.50 $\pm$ 5.56 & 224.16 $\pm$ 22.47 & 0.349 $\pm$ 0.039 \\
SDSSJ1105+0745 & 5268 $\pm$ 262 & 15224 $\pm$ 2513 & 7.83 $\pm$ 0.18 & 16.16 $\pm$ 1.71 & 253.36 $\pm$ 26.83 & 0.026 $\pm$ 0.003 \\
SDSSJ1118+5803 & 4107 $\pm$ 150 & 12589 $\pm$ 2159 & 25.00 $\pm$ 0.74 & 28.88 $\pm$ 3.15 & 275.20 $\pm$ 29.99 & 0.077 $\pm$ 0.011 \\
SDSSJ1122+0117 & 1534 $\pm$ 164 & 7596 $\pm$ 1698 & 4.94 $\pm$ 0.08 & 12.84 $\pm$ 1.31 & 17.08 $\pm$ 1.75 & 0.245 $\pm$ 0.029 \\
SDSSJ1128+1023 & 1244 $\pm$ 150 & 6968 $\pm$ 1366 & 10.94 $\pm$ 0.28 & 19.11 $\pm$ 2.04 & 16.71 $\pm$ 1.79 & 0.555 $\pm$ 0.073 \\
SDSSJ1139+5911 & 2459 $\pm$ 150 & 8101 $\pm$ 1995 & 15.60 $\pm$ 0.14 & 22.81 $\pm$ 2.25 & 77.92 $\pm$ 7.67 & 0.170 $\pm$ 0.018 \\
SDSSJ1141+0241 & 2114 $\pm$ 263 & 8755 $\pm$ 2906 & 3.37 $\pm$ 0.03 & 10.60 $\pm$ 1.04 & 26.76 $\pm$ 2.63 & 0.107 $\pm$ 0.011 \\
SDSSJ1152--0005 & 2693 $\pm$ 150 & 7488 $\pm$ 1375 & 77.88 $\pm$ 2.72 & 50.97 $\pm$ 5.69 & 208.85 $\pm$ 23.31 & 0.316 $\pm$ 0.046 \\
SDSSJ1157--0022 & 5662 $\pm$ 214 & 13711 $\pm$ 2510 & 140.86 $\pm$ 3.26 & 68.55 $\pm$ 7.25 & 1241.78 $\pm$ 131.30 & 0.096 $\pm$ 0.012 \\
SDSSJ1157+0412 & 1522 $\pm$ 150 & 7755 $\pm$ 1910 & 14.61 $\pm$ 0.67 & 22.07 $\pm$ 2.59 & 28.90 $\pm$ 3.38 & 0.428 $\pm$ 0.070 \\
SDSSJ1203+0229 & 2963 $\pm$ 150 & 10524 $\pm$ 1714 & 35.99 $\pm$ 0.24 & 34.65 $\pm$ 3.38 & 171.89 $\pm$ 16.76 & 0.177 $\pm$ 0.018 \\
SDSSJ1223+0240 & 2521 $\pm$ 150 & 7832 $\pm$ 1657 & 8.15 $\pm$ 0.11 & 16.48 $\pm$ 1.66 & 59.21 $\pm$ 5.96 & 0.117 $\pm$ 0.013 \\
SDSSJ1243+0252 & 1168 $\pm$ 150 & 5511 $\pm$ 1239 & 4.31 $\pm$ 0.23 & 11.99 $\pm$ 1.45 & 9.24 $\pm$ 1.12 & 0.395 $\pm$ 0.069 \\
SDSSJ1246+0222 & 2782 $\pm$ 159 & 9034 $\pm$ 1683 & 45.44 $\pm$ 0.32 & 38.93 $\pm$ 3.80 & 170.30 $\pm$ 16.63 & 0.226 $\pm$ 0.024 \\
SDSSJ1300+5641 & 2104 $\pm$ 170 & 9338 $\pm$ 1766 & 13.98 $\pm$ 0.21 & 21.60 $\pm$ 2.19 & 54.01 $\pm$ 5.49 & 0.219 $\pm$ 0.026 \\
SDSSJ1300+6139 & 4644 $\pm$ 239 & 11765 $\pm$ 1589 & 9.25 $\pm$ 0.10 & 17.57 $\pm$ 1.75 & 214.11 $\pm$ 21.31 & 0.037 $\pm$ 0.004 \\
SDSSJ1307--0036 & 2556 $\pm$ 150 & 9077 $\pm$ 1663 & 48.84 $\pm$ 0.53 & 40.36 $\pm$ 4.02 & 149.05 $\pm$ 14.85 & 0.278 $\pm$ 0.031 \\
SDSSJ1307+0107 & 4890 $\pm$ 231 & 15868 $\pm$ 2613 & 183.28 $\pm$ 2.67 & 78.19 $\pm$ 7.93 & 1056.59 $\pm$ 107.19 & 0.147 $\pm$ 0.017 \\
SDSSJ1331+0131 & 1569 $\pm$ 150 & 7615 $\pm$ 1448 & 11.05 $\pm$ 0.44 & 19.20 $\pm$ 2.19 & 26.69 $\pm$ 3.05 & 0.351 $\pm$ 0.054 \\
SDSSJ1341--0053 & 2300 $\pm$ 150 & 8993 $\pm$ 1537 & 32.02 $\pm$ 0.64 & 32.68 $\pm$ 3.40 & 97.72 $\pm$ 10.18 & 0.278 $\pm$ 0.034 \\
SDSSJ1342--0053 & 4301 $\pm$ 150 & 10310 $\pm$ 1888 & 23.61 $\pm$ 0.14 & 28.06 $\pm$ 2.72 & 293.32 $\pm$ 28.47 & 0.068 $\pm$ 0.007 \\
SDSSJ1342+5642 & 2676 $\pm$ 150 & 8311 $\pm$ 1699 & 4.99 $\pm$ 0.07 & 12.90 $\pm$ 1.31 & 52.18 $\pm$ 5.31 & 0.081 $\pm$ 0.009 \\
SDSSJ1343+0004 & 1873 $\pm$ 150 & 7648 $\pm$ 2218 & 25.02 $\pm$ 0.42 & 28.89 $\pm$ 2.96 & 57.24 $\pm$ 5.87 & 0.370 $\pm$ 0.044 \\
SDSSJ1344--0015 & 2086 $\pm$ 150 & 8591 $\pm$ 2372 & 30.36 $\pm$ 0.42 & 31.82 $\pm$ 3.22 & 78.27 $\pm$ 7.92 & 0.329 $\pm$ 0.038 \\
SDSSJ1344+0005 & 5717 $\pm$ 150 & 14506 $\pm$ 2272 & 113.00 $\pm$ 2.11 & 61.40 $\pm$ 6.35 & 1133.83 $\pm$ 117.35 & 0.084 $\pm$ 0.010 \\
SDSSJ1344+4416 & 1372 $\pm$ 150 & 8145 $\pm$ 2017 & 8.30 $\pm$ 0.18 & 16.63 $\pm$ 1.75 & 17.68 $\pm$ 1.86 & 0.398 $\pm$ 0.051 \\
SDSSJ1345--0259 & 3097 $\pm$ 205 & 11282 $\pm$ 1869 & 4.39 $\pm$ 0.03 & 12.10 $\pm$ 1.18 & 65.57 $\pm$ 6.40 & 0.057 $\pm$ 0.006 \\
SDSSJ1349+0204 & 2990 $\pm$ 681 & 9396 $\pm$ 1882 & 2.48 $\pm$ 0.11 & 9.10 $\pm$ 1.05 & 45.98 $\pm$ 5.32 & 0.046 $\pm$ 0.007 \\
SDSSJ1355+6440 & 1527 $\pm$ 150 & 7214 $\pm$ 1614 & 9.14 $\pm$ 0.34 & 17.46 $\pm$ 1.96 & 23.00 $\pm$ 2.59 & 0.337 $\pm$ 0.050 \\
SDSSJ1437+0007 & 2278 $\pm$ 151 & 8801 $\pm$ 2287 & 74.76 $\pm$ 0.79 & 49.94 $\pm$ 4.97 & 146.48 $\pm$ 14.57 & 0.433 $\pm$ 0.048 \\
SDSSJ1505+0342 & 1400 $\pm$ 150 & 7570 $\pm$ 1644 & 7.98 $\pm$ 0.12 & 16.32 $\pm$ 1.66 & 18.08 $\pm$ 1.84 & 0.374 $\pm$ 0.044 \\
\enddata
\end{deluxetable}
\begin{deluxetable}{lcccccc}
\tabletypesize{\tiny}
\tablecolumns{7}
\tablenum{2}
\tablecaption{Line profile and source property estimates}
\tablehead{\colhead{Object Name} & \colhead{FWHM\tablenotemark{a}} & \colhead{FWZI\tablenotemark{a}} &
\colhead{$L_{5100}$\tablenotemark{b}} & \colhead{$R_{\rm BLR}$\tablenotemark{c}} &
\colhead{$M_{\rm BH}$\tablenotemark{d}} & \colhead{$\dot{M} / \dot{M}_{\rm Edd}$}}
\startdata
SDSSJ1510+0058 & 4906 $\pm$ 266 & 11785 $\pm$ 2483 & 21.63 $\pm$ 0.35 & 26.86 $\pm$ 2.75 & 365.27 $\pm$ 37.34 & 0.050 $\pm$ 0.006 \\
SDSSJ1519+0016 & 4387 $\pm$ 171 & 10600 $\pm$ 1802 & 100.02 $\pm$ 1.65 & 57.76 $\pm$ 5.92 & 628.19 $\pm$ 64.34 & 0.135 $\pm$ 0.016 \\
SDSSJ1519+5908 & 1278 $\pm$ 150 & 6936 $\pm$ 1521 & 7.95 $\pm$ 0.30 & 16.29 $\pm$ 1.84 & 15.03 $\pm$ 1.70 & 0.448 $\pm$ 0.067 \\
SDSSJ1535+5754 & 3834 $\pm$ 322 & 10655 $\pm$ 1295 & 12.30 $\pm$ 0.09 & 20.26 $\pm$ 1.98 & 168.21 $\pm$ 16.43 & 0.062 $\pm$ 0.006 \\
SDSSJ1538+4440 & 3242 $\pm$ 188 & 8706 $\pm$ 1439 & 3.79 $\pm$ 0.03 & 11.24 $\pm$ 1.11 & 66.77 $\pm$ 6.59 & 0.048 $\pm$ 0.005 \\
SDSSJ1554+3238 & 5312 $\pm$ 382 & 10818 $\pm$ 1375 & 17.46 $\pm$ 0.29 & 24.14 $\pm$ 2.47 & 384.87 $\pm$ 39.40 & 0.038 $\pm$ 0.005 \\
SDSSJ1613+3717 & 4789 $\pm$ 217 & 11267 $\pm$ 1272 & 9.27 $\pm$ 0.10 & 17.58 $\pm$ 1.75 & 227.90 $\pm$ 22.64 & 0.034 $\pm$ 0.004 \\
SDSSJ1619+4058 & 2436 $\pm$ 150 & 8016 $\pm$ 2087 & 7.88 $\pm$ 0.17 & 16.21 $\pm$ 1.70 & 54.35 $\pm$ 5.71 & 0.123 $\pm$ 0.016 \\
SDSSJ1623+4104 & 1586 $\pm$ 150 & 8242 $\pm$ 1964 & 15.86 $\pm$ 0.37 & 23.00 $\pm$ 2.43 & 32.67 $\pm$ 3.45 & 0.411 $\pm$ 0.053 \\
SDSSJ1654+3925 & 2402 $\pm$ 207 & 7030 $\pm$ 1450 & 3.60 $\pm$ 0.09 & 10.96 $\pm$ 1.17 & 35.71 $\pm$ 3.81 & 0.085 $\pm$ 0.011 \\
SDSSJ1659+6202 & 3990 $\pm$ 150 & 9829 $\pm$ 1132 & 210.04 $\pm$ 2.34 & 83.71 $\pm$ 8.35 & 752.96 $\pm$ 75.10 & 0.236 $\pm$ 0.026 \\
SDSSJ1717+5815 & 3292 $\pm$ 159 & 9966 $\pm$ 1622 & 208.76 $\pm$ 2.27 & 83.45 $\pm$ 8.31 & 511.13 $\pm$ 50.91 & 0.346 $\pm$ 0.038 \\
SDSSJ1719+5937 & 4973 $\pm$ 216 & 20280 $\pm$ 3400 & 137.49 $\pm$ 5.19 & 67.72 $\pm$ 7.66 & 946.25 $\pm$ 106.97 & 0.123 $\pm$ 0.019 \\
SDSSJ1720+5540 & 3469 $\pm$ 244 & 10064 $\pm$ 1750 & 19.96 $\pm$ 0.20 & 25.80 $\pm$ 2.56 & 175.40 $\pm$ 17.40 & 0.096 $\pm$ 0.011 \\
SDSSJ2058--0650 & 2058 $\pm$ 150 & 8580 $\pm$ 1663 & 4.15 $\pm$ 0.10 & 11.77 $\pm$ 1.25 & 28.15 $\pm$ 2.99 & 0.125 $\pm$ 0.016 \\
SDSSJ2349--0036 & 2749 $\pm$ 150 & 8662 $\pm$ 1328 & 2.13 $\pm$ 0.10 & 8.43 $\pm$ 0.99 & 35.98 $\pm$ 4.21 & 0.050 $\pm$ 0.008 \\
SDSSJ2351--0109 & 3062 $\pm$ 153 & 9042 $\pm$ 1614 & 64.75 $\pm$ 0.43 & 46.47 $\pm$ 4.53 & 246.21 $\pm$ 24.00 & 0.223 $\pm$ 0.023 \\
\enddata
\tablenotetext{a}{FWHM and FWZI are measured in ${\rm km\ s^{-1}}$}\\
\tablenotetext{b}{$L_{5100}$ are given in units of $10^{42} {\rm erg\ s^{-1}}$}\\
\tablenotetext{c}{$R_{\rm BLR}$ are in units of $10^{15} {\rm cm}$}\\
\tablenotetext{d}{$M_{\rm BH}$ are expressed in units of $10^5 {\rm M_\odot}$}
\end{deluxetable}
\end{document}